\documentclass[pdftex,twocolumn,epjc3]{svjour3}          
\usepackage{float}
\usepackage{bm}
\usepackage{amsmath}
\usepackage{amssymb}
\usepackage{graphicx}
\usepackage{esint}
\usepackage{multirow}
\usepackage{microtype}

\providecommand{\tabularnewline}{\\}
\allowdisplaybreaks[1]

\RequirePackage[T1]{fontenc}
\RequirePackage{fix-cm}
\smartqed  

\RequirePackage{graphicx}
\RequirePackage{mathptmx}      
\RequirePackage{flushend}
\RequirePackage[numbers,sort&compress]{natbib}
\RequirePackage[colorlinks,citecolor=blue,urlcolor=blue,linkcolor=blue]{hyperref}

\journalname{Eur. Phys. J. A}

\begin{document}

\title{Perturbative quantum Monte Carlo calculation with high-fidelity nuclear forces}

\author{Jun Liu\thanksref{addr1}
        \and
        Teng Wang\thanksref{addr2} 
        \and
        Bing-Nan Lu\thanksref{e1,addr1}
}

\thankstext{e1}{e-mail: bnlv@gscaep.ac.cn (corresponding author)}

\institute{Graduate School of China Academy of Engineering Physics, Beijing 100193, China\label{addr1}
          \and
          School of Physics, Peking University, Beijing 100871, China\label{addr2}
}

\date{Received: date / Accepted: date}

\maketitle

\begin{abstract}
Quantum Monte Carlo (QMC) is a family of powerful tools for addressing
quantum many-body problems. However, its applications are often plagued by the fermionic sign problem. A promising strategy is
to simulate an interaction without sign problem as the zeroth order and treat
the other pieces as perturbations. 
According to this scheme, we construct precision nuclear chiral forces on the lattice and make perturbative calculations around a sign-problem-free interaction respecting the Wigner-SU4 symmetry. 
We employ the recently developed perturbative QMC (ptQMC) method to calculate the perturbative energies up to the second order.
This work presents the first ptQMC calculations for two-body next-to-next-to-next-to leading order (N$^3$LO) chiral forces and elucidates how the hierarchical nature of the chiral interactions helps organize and simplify the ptQMC calculations.
We benchmark the algorithm for the deuteron, where exact solutions serve as rigorous reference points.
We also reproduce the famous Tjon line by correlating the perturbative $^{4}$He binding energies with the non-perturbative $^{3}$H binding energies. 
These comprehensive demonstrations underscore the efficacy of ptQMC in resolving high-fidelity nuclear interactions, establishing its potential as a robust tool for \textit{ab initio} nuclear structure studies.
\end{abstract}

\section{Introduction}

The quantum many-body problem is formidably difficult as the number
of relevant states increases exponentially with the particle number.
One solution is to employ Monte Carlo algorithms to stochastically
sample the enormous Hilbert space, namely the quantum Monte Carlo (QMC)
method. Since the seminal work of Metropolis et al.\cite{metropolis1953equation}, the QMC method has grown into a large family of algorithms
addressing quantum many-body problems in nuclear physics~\cite{PhysRevC.52.718, RevModPhys.87.1067, PPNP63-117, NLEFT}, condensed matter~\cite{PhysRevLett.45.566,RevModPhys.73.33,PhysRevX.3.031010}, ultracold atoms~\cite{PhysRevA.78.023625,PhysRevA.84.061602,PhysRevA.101.063615} and quantum chemistry~\cite{ref11_in_PRL128-242501,ref12_in_PRL128-242501}.
Here we focus on its application for nuclear \textit{ab initio} calculations, which attempt
to solve the many-body Schr{\"o}dinger equation exactly starting from bare nuclear forces.
Such calculations are extremely challenging and have only become feasible in the last few decades.
The QMC algorithms have played an essential role in this process. So far, a variety of nuclear QMC methods have proven their success in vastly
different contexts. On the one hand, several nuclear QMC methods do not rely on a specific basis and solve the problem
directly in the coordinate representation. The examples of this type
include the variational Monte Carlo\cite{NPA361-399}, Green's function
Monte Carlo~\cite{PhysRevC.36.2026,PhysRevC.38.1879}, auxiliary field diffusion
Monte Carlo\cite{PLB446-99} and nuclear lattice effective field theory~\cite{PhysRevC.70.014007,PhysRevC.72.024006}. On the other hand, we have the shell-model basis or momentum basis methods, including the shell model Monte Carlo method\cite{Phys.rep278-1}, Monte
Carlo shell model\cite{PPNP47-319}, determinantal diagrammatic Monte
Carlo~\cite{PhysRevLett.96.160402,New.J.Phys8-153} and Monte Carlo improved many-body perturbation
theory\cite{PhysRevLett.122.042501}. 
Moreover, as a most recent advance, the powerful artificial neutron network has been employed as an efficient variational ansatz for solving the nuclear variational Monte Carlo problems with deep learning techniques~\cite{PhysRevLett.127.022502, PhysRevC.107.034320}.


In most QMC calculations, the problems can be reduced to evaluating
a high-dimensional integral. In some fortunate scenarios, when the integrand
is positive definite, QMC sampling becomes most efficient. In these cases, the computational
complexity of the QMC method is polynomial rather than exponential.
Famous examples include Lattice Quantum Chromodynamics at zero baryon density\cite{ProgThorPhys.110.615},
the repulsive Fermi-Hubbard model at half-filling\cite{PhysRevB.80.075116}
and low-energy nuclear systems in the Wigner SU4 limit~\cite{PhysRev.51.106,PhysRevLett.117.132501,PhysRevLett.127.062501,PLB797-134863}. However, for repulsive
interactions or unbalanced fermions, the amplitudes are usually not
positive-definite or even complex, so we have to handle the integral
of highly oscillating functions or the notorious ``sign problem''.
For severe sign problems, the computational time required increases exponentially
with the system size. Furthermore, it was found that the sign problem
is actually an NP-Hard problem that cannot be solved with a universal algorithm\cite{PhysRevLett.94.170201}.
Nevertheless, for many important systems we can still develop specific
algorithms to alleviate the sign problem based on the special symmetry
or knowledge of the phase space. For example, in the fixed node approximation,
the walkers in Green's function Monte Carlo calculations are confined
to the region of the configuration space with the same sign~\cite{J.Chem.Phys.77-349,PhysRevC.62.014001}.
In the Lefschetz thimble method, the configuration space is complexified and distorted to minimize phase oscillations.~\cite{PhysRevD.86.074506,PhysRevD.88.051501,PhysRevD.88.051502,JHEP1310-147}.
Searching for such sign-problem-free algorithms has been a central and active
topic in QMC research.

Nuclear Lattice Effective Field Theory (NLEFT) is a powerful QMC
algorithm specifically designed for nuclear \textsl{ab initio} calculations.
In this method, the nuclear forces built from the chiral effective
field theory are discretized and regularized with a cubic spatial
lattice; then the corresponding nuclear many-body Schr{\"o}dinger equations
are solved using the auxiliary field Monte Carlo techniques. In recent
years, NLEFT has been successfully applied to address various
nuclear physics problems from the first principles, \textit{e.g.}, solving
the nuclear ground states~\cite{EPJA31-105,PhysRevLett.104.142501,EPJA45-335,PLB732-110,PLB797-134863,PhysRevLett.128.242501,Nature630-59} and excited
states~\cite{PhysRevLett.112.102501,Nat.Comm.14-2777,PhysRevLett.132.062501,arxiV2411.14935},
exploring the nuclear instrinsic densities and clustering phenomena~\cite{PhysRevLett.106.192501,PhysRevLett.109.252501,PhysRevLett.110.112502,PhysRevLett.119.222505,arxiV2411.17462}, extracting
the nucleus-nucleus scattering cross sections~\cite{PhysRevC.86.034003,Nature528-111}, simulating the zero and finite temperature nuclear
matter~\cite{PhysRevLett.117.132501,PhysRevLett.125.192502,PLB850-138463,PhysRevLett.132.232502}
and calculating the structure of the hypernucleus~\cite{PhysRevLett.115.185301,EPJA56-24,EPJA60-215}. 
Despite these successes, the NLEFT
calculations are also hampered by the sign problem as with other QMC
methods. For instance, the repulsive Coulomb force can induce an explicit
imaginary phase factor in the auxiliary field transformation, thus
cannot be simulated non-perturbatively~\cite{PLB797-134863,Nature630-59}.
This also occurs for general many-body forces. Due to the non-commutative
Pauli matrices, the spin- and isospin-dependent terms result in strong sign problem\cite{EPJA51-92}. In generalizations
of NLEFT, such as the pinhole algorithm for intrinsic densities\cite{PhysRevLett.119.222505}
and the pinhole trace algorithm for thermodynamics\cite{PhysRevLett.125.192502},
the unpaired pinholes can be the origin of the sign oscillation; thus,
for low-temperature or large projection time, the statistical errors
increase rapidly. Consequently, the sign problem largely limits
the available interactions and the parameter space that can be explored
since the early stage of the NLEFT~\cite{EPJA31-105,EPJA35-343,EPJA41-125,PhysRevLett.104.142501,EPJA45-335,PLB732-110}. Until now, in most
NLEFT calculations, only the leading order interactions can be simulated
non-perturbatively, while higher order operators are incoorperated
with the first order perturbation theory~\cite{PhysRevLett.106.192501,PhysRevLett.109.252501,PhysRevLett.110.112502,PhysRevLett.112.102501}. Such a scheme was recently improved
with the wave function matching method, in which a local unitary transformation
is applied to enhance the precision of the first order perturbative
calculations. It was shown that the experimental nuclear binding energies up to
$A=56$ can be well reproduced with a perturbative expansion around
a simple leading order Hamiltonian\cite{Nature630-59}.

The leading order or non-perturbative Hamiltonian in the NLEFT calculations can
be chosen arbitrarily as long as the corresponding sign problem is
under control. A suitable choice is a central force with the Wigner
SU4 symmetry\cite{PhysRev.51.106}, which means that the interaction is
independent of both the spin and the isospin. The Wigner SU4 symmetry
is an approximate symmetry of the nuclear force and can be exploited
to construct a sign-problem-free Hamiltonian~\cite{PhysRevLett.118.202501,EPJA56-113,Few_Body_Syst_58_26}. In Ref.\cite{PhysRevLett.117.132501} it was discovered
that the locality of the nuclear force is indispensable for describing the nuclear clustering. Later the leading order three-body force
was also included as an essential element of nuclear binding\cite{PLB797-134863}.
Such SU4 interactions give fairly well reproduction of the experimental
binding energies, charge radii, as well as the nuclear matter properties
and have been applied in studying the nuclear liquid-gas phase transition\cite{PhysRevLett.125.192502},
clustering at finite temperatures\cite{PLB850-138463}, monopole transition
in alpha particle\cite{PhysRevLett.132.062501} and tomography of the low-lying
states of carbon isotopes\cite{Nat.Comm.14-2777}. Starting from
such effective interactions, we expect that most of the essential
part of the nuclear force has already been absorbed, and the complicated
spin- or isospin-dependent components and many-body forces only play minor
roles and can be treated safely with perturbation theory.

To improve the precision of the NLEFT calculations, it is desirable
to consider the higher order perturbative corrections. However, in
conventional Rayleigh-Schr{\"o}dinger perturbation theory the second order
perturbative energy consists of contributions from all excited energy
levels, which is difficult to obtain in projection Monte Carlo methods that only 
target the lowest levels. To overcome this challenge, in Ref.~\cite{PhysRevLett.128.242501}
a perturbative quantum Monte Carlo (ptQMC) method was proposed to
calculate the second order perturbative energies efficiently within
the auxiliary field formalism. The key idea is to represent the perturbative
energies using partial amplitudes that can be calculated precisely
with Monte Carlo sampling. To demonstrate the precision of the method,
in Ref.\cite{PhysRevLett.128.242501} a nuclear next-to-next-to leading order
(N$^{2}$LO) chiral force was constructed and solved by expanding
the Hamiltonian around a simple SU4 interaction. The resulting binding energies
of $^{3}$H, $^{4}$He and $^{16}$O are mostly independent of the strength of the SU4
interaction and agree well with the experimental values. Later
the second order perturbative calculations were also implemented in
the diffusion Monte Carlo method~\cite{PhysRevResearch.5.L042021} and
applied to incorporate the nonlocal interactions~\cite{PhysRevC.111.015801}, which was previously
difficult to simulate in the continuum QMC methods.

The effort of perturbatively solving the NLEFT can be continued in two
directions. We can either consider even-higher order perturbative
corrections or include more precision nuclear forces, both of which
are highly motivated and challenging. On one hand, the perturbative
series may converge slowly for shallow bound states near the continuum,
which requires third or higher order corrections. On the other hand,
the higher order chiral forces with complex spin or isospin dependencies may play essential roles in certain environment, \textit{e.g.}, the short-range correlations induced by the complex tensor forces are important in the electroweak processes~\cite{PhysRevLett.127.062501}. 
Thus it is attractive
to solve the high-fidelity chiral forces with the ptQMC method and
check their impacts in many-body systems. Both directions require
proper developments of the interactions together with substantial
improvements of the algorithms. Here we choose to firstly improve
the chiral forces to the next-to-next-to-next-to leading order (N$^{3}$LO)
and focus on the deeply bound nuclei such as $^{4}$He,
for which the second order perturbation theory is sufficient for the
desired precision. As a demonstration of our method, here we constrain ourselves to the two-body chiral forces only.
We add the 15 extra two-body contact terms at N$^{3}$LO~\cite{NPA361-399,PhysRevC.68.041001} and
 omit all the three-body forces as well as long-range pion-exchange diagrams except for the
one-pion-exchange potential. 
Note that there is no difficulty in applying our method to incorporate all these complicated interactions.
A detailed analysis based on the full N$^3$LO chiral force will be implemented in the near future.

This paper is organized as follows. In Sect.II we present the details
of the lattice SU4 interaction and the N$^{3}$LO chiral force, then
briefly discuss the ptQMC method. In Sect.III, we present numerical
results for $^{2}$H and $^{4}$He. Finally in Sect.IV we
close with a summary and perspective.

\section{Theoretical framework}

\subsection{Lattice SU4 interaction}

In this section, we present the details of the SU4 interaction to be
solved non-perturbatively. We refer to the two-body part of the pionless
effective field theory presented in Ref.~\cite{PLB797-134863}.
On a periodic $L^{3}$ cube with lattice coordinates $\bm{n}=(n_{x,}n_{y},n_{z})$,
the Hamiltonian is 
\begin{equation}
\setlength\abovedisplayskip{18pt}
\setlength\belowdisplayskip{18pt}
H_{0}=K+\frac{1}{2}C_{{\rm SU4}}\sum_{\bm{n}}:\tilde{\rho}^{2}(\bm{n}):,\label{eq:HSU4}
\end{equation}
where $K$ is the kinetic energy term with nucleon mass $m=938.9$
MeV and the $::$ symbol indicates normal ordering. In what follows,
we fix the lattice spacing to $a=1.0$ fm and use the unit system
$\hbar=c=a=1$.
The
locally smeared density operator $\tilde{\rho}(\bm{n})$ in Eq.~(\ref{eq:HSU4})
is defined as 
\begin{equation}
\setlength\abovedisplayskip{18pt}
\setlength\belowdisplayskip{18pt}
\tilde{\rho}(\bm{n})=\sum_{i}\tilde{a}_{i}^{\dagger}(\bm{n})\tilde{a}_{i}(\bm{n})+s_{L}\sum_{|\bm{n}^{\prime}-\bm{n}|=1}\sum_{i}\tilde{a}_{i}^{\dagger}(\bm{n}^{\prime})\tilde{a}_{i}(\bm{n}^{\prime}),
\label{eq:localsmearing}
\end{equation}
where $i$ is the joint spin-isospin index. The non-locally smear-\\ed annihilation and creation operators are defined as 
\begin{equation}
\setlength\abovedisplayskip{18pt}
\setlength\belowdisplayskip{18pt}
\tilde{a}_{i}(\bm{n})=a_{i}(\bm{n})+\sum_{\bm{n}^{\prime}}f(|\bm{n}^{\prime}-\bm{n}|)a_{i}(\bm{n}^{\prime}),
\label{eq:nonlocalsmearing}
\end{equation}
where $f(x)$ is a smearing function with a finite width. In this work, we implement the non-local smearing using the fast Fourier transform (FFT). We transform the single-particle wave functions into the momentum space, multiply them with a Gaussian cutoff function
\begin{equation}
\setlength\abovedisplayskip{18pt}
\setlength\belowdisplayskip{18pt}
\tilde{f}(p)=\exp\left(-p^{2}/(2\Lambda_{{\rm SU4}}^{2})\right)
\end{equation}
and transform them back to the coordinate space. Here $\tilde{f}(p)$
is the Fourier transform of $f(x)$. Note that in Ref.\cite{PLB797-134863}
we use a lattice spacing $a=1.32$ fm and implement the non-local
smearing by summing over the nearest neighboring lattice sites.
For a lattice spacing as small as $a=1.0$ fm, the contributions from the
next-to-nearest neighboring or even further lattice sites are essential
and can be taken into account with a smooth Gaussian smearing function.
The summation over the spin and isospin implies that the interaction
is SU4 invariant. The parameter $s_{L}$ controls the strength of
the local part of the interaction. Both local and non-local smearing
have an impact on the range of the interaction. The parameters $C_{{\rm SU4}}$
give the strength of the two-body interactions. In this work, we use
the parameter set $C_{{\rm SU4}}=-3.79\times10^{-5}$ MeV$^{-2}$,
$s_{L}=0.182$ and $\Lambda_{{\rm SU4}}=300$ MeV. Such a simple interaction
together with a properly chosen three-body force can reproduce the
binding energies and charge density distributions of light nuclei
from $^{3}$H to the Ca isotopes as exemplified in Ref.\cite{PLB797-134863}. 

\begin{figure}[h]
\begin{centering}
\includegraphics[width=0.4\textwidth]{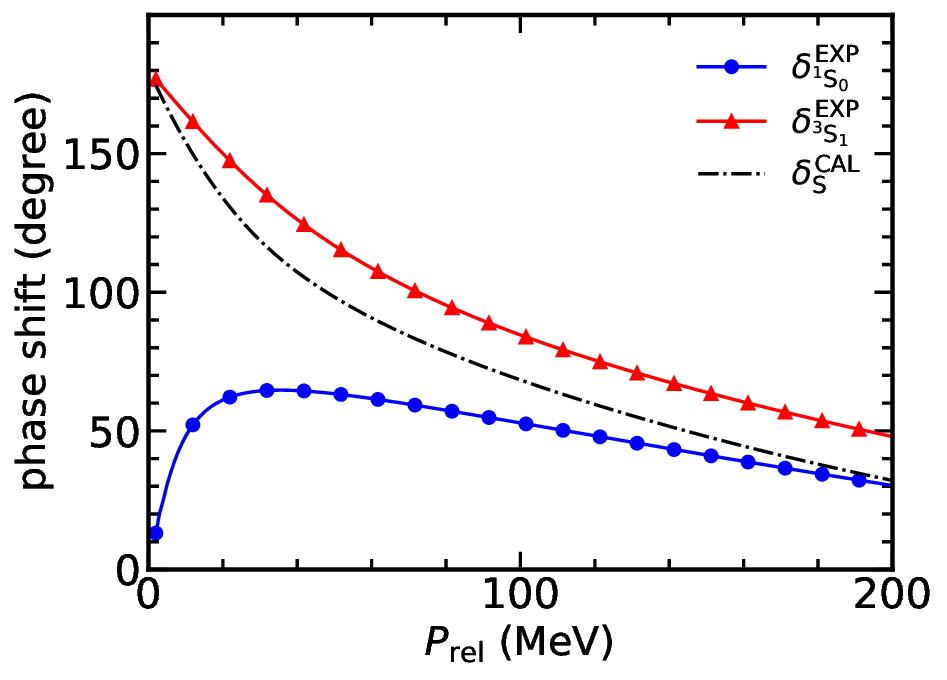}
\par\end{centering}
\caption{\label{fig:zeroth_phase_shifts}The triangles (red) and circles (blue)
denote the empirical $^{1}S_{0}$ and $^{3}S_{1}$ phase shifts, respectively.
The dash-dotted curve represents the results from the SU4 Hamiltonian
Eq.~(\ref{eq:HSU4}).}
\end{figure}

In Fig.~\ref{fig:zeroth_phase_shifts} we show the neutron-proton $S$-wave phase
shifts calculated with Eq.~(\ref{eq:HSU4}) (dash-dotted line) compared to the empirical $^{1}S_{0}$ (circles) and $^{3}S_{1}$ (triangles) phase shifts. It is well known that the nuclear force is close to a unitary limit, \textit{i.e.}, the $S$-wave scattering lengths $a_{^{1}S_{0}}=-23.7$
fm and $a_{^{3}S_{1}}=5.425$ fm are much larger than the range of the nuclear force $r_{{\rm eff}}\approx m_{\pi}^{-1}\approx1$
fm. The SU4 Hamiltonian Eq.~(\ref{eq:HSU4}) gives a scattering length
$a_{0}=9.2$ fm and the phase shifts are approximately the average
of the two $S$-wave channels. The corresponding effective range $r_{0}=2.2$
fm is also at the middle of the two empirical values $r_{^{1}S_{0}}=2.77$
fm and $r_{^{3}S_{1}}=1.75$ fm. In reality, due to the spin-isospin dependent
contact terms and the long-range tenor forces in the realistic nuclear
forces, the $^{1}S_{0}$ and $^{3}S_{1}$ energy levels split, and we
obtain a shallow bound state at $E(^2$H$)=-2.22$~MeV. 


\subsection{Lattice N$^{3}$LO chiral interaction}

In this section, we present the details of the N$^{3}$LO lattice chiral
interaction. 
The construction of the chiral forces on the lattice follows the same procedure as in the continuum, with slight adaptations of the operator basis to Monte Carlo simulations.
We fix the low-energy constants by fitting to the empirical phase shifts and mixing angles~\cite{PhysRevC.98.044002} based on the improved spherical wall method~\cite{Phys.Lett.B760.309}.
For now, there exist several high-precision lattice chiral forces amenable to Monte Carlo simulations.
The first N$^2$LO~\cite{EPJA53-83} and N$^3$LO~\cite{PhysRevC.98.044002} lattice chiral forces were built by substituting the spatial derivatives by finite differences.
In Ref.~\cite{PhysRevLett.128.242501} we construct a N$^2$LO lattice chiral force with the various momenta implemented by the FFT algorithm, which facilitates the second-order perturbative calculations.
Recently, a rank-one-operator technique was proposed to solve finite-temperature nuclear matter with perturbation theory, in which the operator basis are deliberately designed for convenience~\cite{PhysRevLett.132.232502}.
In the wave function matching method, Galilean invariance restoration terms and higher-order three-body forces were introduced to improve the capability of the lattice chiral forces in describing finite nuclei~\cite{Nature630-59}.
We emphasize that all these NN interactions reproduce the neutron-proton phase shifts equally well. 
This work can be viewed as an extension of Ref.~\cite{PhysRevLett.128.242501}, thus we build the lattice chiral forces in the same way.
We first write down the contact terms as well as the pion-exchange potentials as functions of the incoming and outgoing momenta, then realize these operators on the lattice using the FFT algorithm. 
The advantage of this protocol is that we can apply the familiar conclusions drawn in the continuum space.
For example, we can use the operators consisting of the momentum transfers and relate the corresponding LECs to the spectroscopic LECs defined for each partial wave with simple linear transformations~\cite{PA747-362, PhysRevC.68.041001}.


Next we explain the detail of the N$^3$LO lattice chiral forces.
We first consider nucleon-nucleon scattering. We denote the
two incoming and two outgoing momenta as $\bm{p}_{1}$, $\bm{p}_{2}$,
$\bm{p}_{1}^{\prime}$ and $\bm{p}_{2}^{\prime}$, respectively. In
terms of the relative momenta $\bm{p}=\frac{1}{2}(\bm{p}_{1}-\bm{p}_{2})$
and $\bm{p}^{\prime}=\frac{1}{2}(\bm{p}_{1}^{\prime}-\bm{p}_{2}^{\prime})$,
we can write the momentum transfer in direct and exchange channels
as
\begin{align*}
\setlength\abovedisplayskip{18pt}
\setlength\belowdisplayskip{18pt}
\bm{q} & =\bm{p}^{\prime}-\bm{p}=\bm{p}_{1}^{\prime}-\bm{p}_{1}=-(\bm{p}_{2}^{\prime}-\bm{p}_{2}),\\
\bm{k} & =\frac{1}{2}(\bm{p}^{\prime}+\bm{p})=\frac{1}{2}(\bm{p}_{1}-\bm{p}_{2}^{\prime})=-\frac{1}{2}(\bm{p}_{2}-\bm{p}_{1}^{\prime}),
\end{align*}
where we have used the momentum conservation $\bm{p}_{1}+\bm{p}_{2}=\bm{p}_{1}^{\prime}+\bm{p}_{2}^{\prime}$.
Because $\bm{k}=(\bm{p}_{1}-\bm{p}_{2}^{\prime})/2$ consists of the momenta
of two different particles, it is involved to implement the
$\bm{k}$-dependent operators on the lattice without breaking
the fundamental symmetries such as Galilean invariance. To resolve
this issue, we follow the prescription of Ref.~\cite{PhysRevLett.111.032501}
to adopt an operator basis with minimal $\bm{k}$-dependence.
This can be achieved by introducing terms proportional
to $\bm{\tau}_{1}\cdot\bm{\tau}_{2}$ with $\rm{\tau}_{1,2}$ the isospin Pauli matrices.

We already know that the choice of the operator basis is not unique due to the antisymmetrization nature of the fermion wave functions, thus we can select the operators that are convenient for the lattice simulations. 
Note that such freedom for choosing the operators was also exploited in the Green's function Monte Carlo calculations~\cite{PhysRevLett.111.032501}, for which the non-local operators are more difficult to simulate than the local ones.
These $\rm{\tau}$-dependent operators span the same subspace
as the isospin-independent operator basis that has been routinely used in the chiral
EFT literature~\cite{RevModPhys.81.1773, Frontiers_in_physics8-98, PhysRep503-1}. The linear transformation between these two basis
can be easily derived by considering the exchange symmetries of the
momenta and indices of the Pauli matrices. In Eq.~(\ref{N3LOcontactoperators}) we present the 24
linearly independent contact terms up to the N$^{3}$LO, where
 $\bm{\sigma}_{1}$ and $\bm{\sigma}_{2}$
are spin matrices, $\bm{\tau}_{1}$ and $\bm{\tau}_{2}$ are isospin
matrices, $B_{1-2}$, $C_{1-7}$ and $D_{1-15}$ are LECs to be determined.
To reduce the lattice artifacts we have introduced an extra soft
single-particle regulator with a momentum cutoff $\Lambda$. In this
work we use $\Lambda=300, 350$ and $400$ MeV, which are all much lower than
the lattice momentum cutoff $\Lambda_{{\rm lat}}=\pi/a\approx600$
MeV. In this way the low-energy physics is solely determined by the
form of the soft regulator, with the unpleasant lattice artifacts
such as the rotational symmetry breaking effects screened and minimized.
In the center of mass frame we have $\bm{p}_{1}=-\bm{p}_{2}$, such
a single-particle regulator is equivalent to the non-local regulator
used in the conventional chiral EFT constructions. However, for moving
frames or particle number greater than three, the single-particle
regulator explicitly breaks the Galilean invariance. The degree of
this symmetry breaking effect decreases as $\mathcal{O}(\Lambda^{-2})$
and should be compensated with the Galilean invariance restoration
terms~\cite{PhysRevC.99.064001, Nature630-59}. We note that the conventional Galilean invariant regulator
that regulates the relative momenta $\bm{p}$ and $\bm{p}^{\prime}$
is more difficult to implement on the lattice and will give different
predictions compared with the lattice single-particle regulator used here.
Furthermore, the operators in Eq.~(\ref{N3LOcontactoperators})
can be easily written as products of one-body density operators in
the coordinate space and decomposed using the auxiliary field transformations,
which is mandatory for a full non-perturbative calculation or a second-order ptQMC calculation.

\begin{eqnarray}
V_{{\rm 2N}}^{(0)}&=&\biggl[B_{1}+B_{2}(\bm{\sigma}_{1}\cdot\bm{\sigma}_{2})\biggr]\exp\left[-\sum_{i=1}^{2}(p_{i}^{6}+p_{i}^{\prime6})/(2\Lambda^{6})\right],\nonumber \\
V_{{\rm 2N}}^{(2)}&=&\biggl[C_{1}q^{2}+C_{2}q^{2}(\bm{\tau}_{1}\cdot\bm{\tau}_{2})+C_{3}q^{2}(\bm{\sigma}_{1}\cdot\bm{\sigma}_{2})\nonumber\\
& &+C_{4}q^{2}(\bm{\sigma}_{1}\cdot\bm{\sigma}_{2})(\bm{\tau}_{1}\cdot\bm{\tau}_{2})+C_{5}\frac{i}{2}(\bm{q}\times\bm{k})\cdot(\bm{\sigma}_{1}+\bm{\sigma}_{2})\nonumber \\
 &  & +C_{6}(\bm{\sigma}_{1}\cdot\bm{q})(\bm{\sigma}_{2}\cdot\bm{q})+C_{7}(\bm{\sigma}_{1}\cdot\bm{q})(\bm{\sigma}_{2}\cdot\bm{q})(\bm{\tau}_{1}\cdot\bm{\tau}_{2})\biggr]\nonumber\\
 & &\times\exp\left[-\sum_{i=1}^{2}(p_{i}^{6}+p_{i}^{\prime6})/(2\Lambda^{6})\right],\nonumber \\
V_{{\rm 2N}}^{(4)} & = & \Biggl[D_{1}q^{4}+D_{2}q^{4}(\bm{\tau}_{1}\cdot\bm{\tau}_{2})+D_{3}q^{4}(\bm{\sigma}_{1}\cdot\bm{\sigma}_{2})\nonumber\\
& &+D_{4}q^{4}(\bm{\sigma}_{1}\cdot\bm{\sigma}_{2})(\bm{\tau}_{1}\cdot\bm{\tau}_{2})+D_{5}q^{2}(\bm{\sigma}_{1}\cdot\bm{q})(\bm{\sigma}_{2}\cdot\bm{q})\nonumber \\
 &  & +D_{6}q^{2}(\bm{\sigma}_{1}\cdot\bm{q})(\bm{\sigma}_{2}\cdot\bm{q})(\bm{\tau}_{1}\cdot\bm{\tau}_{2})+D_{7}q^{2}k^{2}\nonumber\\
 & &+D_{8}q^{2}k^{2}(\bm{\sigma}_{1}\cdot\bm{\sigma}_{2})+D_{9}(\bm{q}\cdot\bm{k})^{2}\nonumber\\
 & &+D_{10}(\bm{q}\cdot\bm{k})^{2}(\bm{\sigma}_{1}\cdot\bm{\sigma}_{2})\nonumber \\
 &  & +D_{11}\frac{i}{2}q^{2}(\bm{q}\times\bm{k})\cdot(\bm{\sigma}_{1}+\bm{\sigma}_{2})\nonumber\\
 & &+D_{12}\frac{i}{2}q^{2}(\bm{q}\times\bm{k})\cdot(\bm{\sigma}_{1}+\bm{\sigma}_{2})(\bm{\tau}_{1}\cdot\bm{\tau}_{2})\nonumber\\
 & &+D_{13}k^{2}(\bm{\sigma}_{1}\cdot\bm{q})(\bm{\sigma}_{2}\cdot\bm{q})\nonumber \\
 &  & +D_{14}k^{2}(\bm{\sigma}_{1}\cdot\bm{q})(\bm{\sigma}_{2}\cdot\bm{q})(\bm{\tau}_{1}\cdot\bm{\tau}_{2})\nonumber\\
 & &+D_{15}\left[(\bm{q}\times\bm{k})\cdot\bm{\sigma}_{1}\right]\left[(\bm{q}\times\bm{k})\cdot\bm{\sigma}_{2}\right]\Biggr]\nonumber\\
 & &\times\exp\left[-\sum_{i=1}^{2}(p_{i}^{6}+p_{i}^{\prime6})/(2\Lambda^{6})\right].\label{N3LOcontactoperators}
\end{eqnarray}

Besides the short-range contact terms, we also need a long-range one-pion-exchange-potential
(OPEP). Recently, a semi-local momentum space regularized chiral potential
was developed up to fifth order\cite{EPJA54.86}. This regularization
method is more convenient than other choices for lattice simulations.
In momentum space we have
\begin{equation}
\setlength\abovedisplayskip{18pt}
\setlength\belowdisplayskip{18pt}
V_{{\rm 1\pi}}=-\frac{g_{A}^{2}f_{\pi}(q^{2})}{4F_{\pi}^{2}}\left[\frac{(\bm{\sigma}_{1}\cdot\bm{q})(\bm{\sigma}_{2}\cdot\bm{q})}{q^{2}+M_{\pi}^{2}}+C_{\pi}^{\prime}\bm{\sigma}_{1}\cdot\bm{\sigma}_{2}\right](\bm{\tau}_{1}\cdot\bm{\tau}_{2}),\label{eq:OPEP_in_P_space}
\end{equation}
with $g_{A}=1.287$ the axial-vector coupling constant, $F_{\pi}=92.2$
MeV the pion decay constant and $M_{\pi}=134.98$ MeV the pion mass.
$f_{\pi}(q^{2})=e^{-(q^{2}+M_{\pi}^{2})/\Lambda_{\pi}^{2}}$ is a
local exponential regulator with $\Lambda_{\pi}=300$ MeV. The constant
$C_{\pi}^{\prime}$ is defined as
\begin{align*}
\setlength\abovedisplayskip{18pt}
\setlength\belowdisplayskip{18pt}
C_{\pi}^{\prime} & =-\left[\frac{\Lambda_{\pi}^{2}-2M_{\pi}^{2}}{3\Lambda_{\pi}^{2}}+2\sqrt{\pi}\frac{M_{\pi}^{3}}{3\Lambda_{\pi}^{3}}\exp(\frac{M_{\pi}^{2}}{\Lambda_{\pi}^{2}}){\rm erfc}(\frac{M_{\pi}}{\Lambda_{\pi}})\right]
\end{align*}
The term proportional to $C_{\pi}^{\prime}$ is a counter term introduced
to remove the short-range singularity from the OPEP\cite{EPJA54.86}. 


The isospin symmetry is not strictly conserved because of the mass difference
between the up and down quarks, as well as the electromagnetic effects.
At leading order this is reflected by the observed differences among
the empirical proton-neutron, proton-proton, and neutron-neutron scattering
parameters. To address this issue we add two short-range charge symmetry
breaking operators to the Hamiltonian,
\begin{align}
\setlength\abovedisplayskip{18pt}
\setlength\belowdisplayskip{18pt}
V_{{\rm CSB}}^{{\rm pp}} & =c_{{\rm pp}}\left(\frac{1+\tau_{1z}}{2}\right)\left(\frac{1+\tau_{2z}}{2}\right)e^{-\sum_{i=1}^{2}(p_{i}^{6}+p_{i}^{\prime6})/(2\Lambda^{6})},\nonumber \\
V_{{\rm CSB}}^{{\rm nn}} & =c_{{\rm nn}}\left(\frac{1-\tau_{1z}}{2}\right)\left(\frac{1-\tau_{2z}}{2}\right)e^{-\sum_{i=1}^{2}(p_{i}^{6}+p_{i}^{\prime6})/(2\Lambda^{6})},\label{eq:VCSB}
\end{align}
where the LECs $c_{{\rm pp,nn}}$ can be determined by fitting to
the empirical proton-proton and neutron-neutron scattering lengths,
respectively. These operators are regulated using the same single-particle
regulators as is used in Eq.~(\ref{N3LOcontactoperators}).

As mentioned above, the single-particle regulator for contact terms
explicitly breaks the Galilean invariance. Although such effects are
from the regulators rather than the interaction itself and thus suppressed
for large $\Lambda$, for a high-precision calculation with the N$^{3}$LO
chiral interaction, it is both consistent and desirable to include
the leading order Galilean invariance restoration (GIR) terms,
\begin{equation}
\setlength\abovedisplayskip{18pt}
\setlength\belowdisplayskip{18pt}
V_{{\rm GIR}}=\left[g_{1}Q^{2}+g_{2}Q^{2}(\bm{\sigma}_{1}\cdot\bm{\sigma}_{2})\right]e^{-\sum_{i=1}^{2}(p_{i}^{6}+p_{i}^{\prime6})/(2\Lambda^{6})},\label{eq:VGIR}
\end{equation}
where $\bm{Q}=\bm{p}_{1}+\bm{p}_{2}=\bm{p}_{1}^{\prime}+\bm{p}_{2}^{\prime}$
is the center of mass momentum. Note that $V_{{\rm GIR}}$ is short-ranged
and need to be regulated as the two-body contact terms in Eq.~(\ref{N3LOcontactoperators}).
The coefficient $g_{1,2}$ can be adjusted to ensure that the two
$S$-wave scattering lengths are independent of the center of mass
momentum $Q$ up to $\mathcal{O}(Q^{2})$. In this work we include $V_{{\rm GIR}}$ for calculations with particle number $A>2$.

Besides the nuclear force, we also include a static Coulo-\\mb force for
protons. In momentum space we have
\begin{equation}
\setlength\abovedisplayskip{18pt}
\setlength\belowdisplayskip{18pt}
V_{{\rm cou}}=\frac{\alpha}{q^{2}}\exp\left(-\frac{q^{2}}{2\Lambda_{{\rm cou}}^{2}}\right)\left(\frac{1+\tau_{1z}}{2}\right)\left(\frac{1+\tau_{2z}}{2}\right),\label{eq:VCOU}
\end{equation}
where $\alpha=1/137$ is the fine structure constant, $\Lambda_{{\rm cou}}=300$
MeV is a momentum cutoff introduced to remove the singularity at $r=0$.
In coordinate space, the potential can be expressed using the error
function,
\[
\tilde{V}_{{\rm cou}}^{{\rm pp}}(r)=\alpha\frac{{\rm erf}(\Lambda_{{\rm cou}}r/2)}{r}.
\]
We have $\tilde{V}_{{\rm cou}}^{{\rm pp}}(0)=\alpha\Lambda_{{\rm cou}}/\sqrt{\pi}$
and $\tilde{V}_{{\rm cou}}^{{\rm pp}}(r)\rightarrow\alpha/r$ for
$\Lambda_{{\rm cou}}\rightarrow\infty$. The potential $\tilde{V}_{{\rm cou}}^{{\rm pp}}$
is smooth for any $r$ and can be implemented on the lattice without any difficulty. We found that the calculated
Coulomb energy is almost independent of $\Lambda_{{\rm cou}}$ for
$\Lambda_{{\rm cou}}\geq300$ MeV and expect that the remaining cutoff
dependencies can be absorbed by the charge symmetry breaking terms
introduced above.

Compared with the high-precision chiral forces construct-\\ed in the
continuum, our lattice version omits the long-range contribution from
the higher-order one-pion and two-pion exchange diagrams. 
These
choices can be partly justified by considering the momentum scales.
In this work we fit the LECs using NN scattering data below 200 MeV
and only consider the light nuclei up to $^{4}$He, thus the typical momentum
scale is much lower than twice the pion mass. For such low momenta,
the many-pion exchange potentials are analytical functions that can be expanded as polynomials of
the momentum transfer and absorbed into the contact terms.
Although there is no difficulty in implementing these long-range components
, in this work we focus mainly on the demonstration
of the method and leave the detailed analysis of the interactions
for future work.

Now we discuss the implementation of the continuum interactions Eq.~(\ref{N3LOcontactoperators}-\ref{eq:VGIR})
on the lattice. Note that the short-ranged interactions $V_{{\rm 2N}}^{(0,2,4)}$,
$V_{{\rm CSB}}$ and $V_{{\rm GIR}}$
are all regulated with the non-local regulators and the long-ranged
interactions $V_{{\rm cou}}$ and $V_{1\pi}$ are both regulated with
the local regulators. All momentum cutoffs appeared in these regulators
are much lower than the lattice cutoff $\Lambda_{{\rm lat}}\approx600$
MeV. As for the case of $V_{{\rm 2N}}$, all these operators can be
safely simulated the lattice with the various momenta calculated using
the FFT. For a fixed value of the cutoff
$\Lambda$, the lattice artifacts are suppressed as powers of $\Lambda/\Lambda_{{\rm lat}}$
and completely vanish in the continuum limit $a\rightarrow0$ or $\Lambda_{{\rm lat}}\rightarrow\infty$.
In this sense, only the physical cutoff $\Lambda$ appearing in the potentials
is of relevance, while the lattice is only employed as a numerical tool
for calculating the integrals. This scenario is different
from the case of the lattice QCD, where the lattice plays a dual role
of the numerical tool as well as a physical regulator.

As requested by the ptQMC method, we need to perform auxiliary
field transformations for every term appeared in the target Hamiltonian~\cite{PhysRevLett.128.242501}.
This could be tedious for the N$^{3}$LO chiral force, especially
for the $Q^{4}$ operators with complicated momentum and spin dependencies.
Fortunately, as will be demonstrated in following sections, these
operators only play a minor role and an improved first order perturbative
calculation is sufficient for the desired precision. As an example, here we only
show the auxiliary-field decomposition of the potentials $V_{{\rm 2N}}^{(0)}$, $V_{{\rm 2N}}^{(2)}$,
and $V_{1\pi}$. The other terms $V_{\rm CSB}$, $V_{\rm GIR}$ and $V_{\rm cou}$ can be transformed
similarly.

The interactions can be written in terms of the following density
and current operators and their partial derivatives,
\begin{align}
\setlength\abovedisplayskip{18pt}
\setlength\belowdisplayskip{18pt}
\bar{\rho} & =\bar{\psi}^{\dagger}\bar{\psi},\qquad\bar{\rho}_{i}=\bar{\psi}^{\dagger}\sigma_{i}\bar{\psi},\nonumber \\
\bar{\rho}_{a} & =\bar{\psi}^{\dagger}\tau_{a}\bar{\psi},\qquad\bar{\rho}_{ia}=\bar{\psi}^{\dagger}\sigma_{i}\tau_{a}\bar{\psi},\nonumber \\
\bar{\theta}_{0,i} & =-\frac{i}{2}\left(\bar{\psi}^{\dagger}\partial_{i}\bar{\psi}-\partial_{i}\bar{\psi}^{\dagger}\bar{\psi}\right),\nonumber \\
\bar{\theta}_{k,i} & =-\frac{i}{2}\left(\bar{\psi}^{\dagger}\sigma_{k}\partial_{i}\bar{\psi}-\partial_{i}\bar{\psi}^{\dagger}\sigma_{k}\bar{\psi}\right).\label{eq:densities_and_currents}
\end{align}
In what follows we use the symbols $i$, $j$ and $k$ for spin indices
and $a$, $b$, $c$ for isospin indices. The indices after the comma
denote the spatial derivatives. The overlined symbols $\bar{\psi}^{\dagger}$
and $\bar{\psi}$ denote the creation and annilation operators smeared
with the non-local regulators introduced in Eq.~(\ref{N3LOcontactoperators}), 
\begin{equation}
\setlength\abovedisplayskip{18pt}
\setlength\belowdisplayskip{18pt}
\bar{\psi}^{\dagger}(\bm{\bm{n}})=\sum_{\bm{n}^{\prime}}g(|\bm{n}-\bm{n}^{\prime}|)\psi^{\dagger}(\bm{n}^{\prime}),
\end{equation}
with $g(r)$ the Fourier transform
\begin{equation}
\setlength\abovedisplayskip{18pt}
\setlength\belowdisplayskip{18pt}
g(\bm{r})=\int\frac{d^{3}\bm{p}}{(2\pi)^{3}}\exp(-p^{6}/2\Lambda^{6})\exp(i\bm{p}\cdot\bm{r}).
\end{equation}
For building the long-range OPEP we use the point density operators
such as $\rho_{ia}=\psi^{\dagger}\sigma_{i}\tau_{a}\psi$. The coordinate
space expressions after second quantization are
\begin{eqnarray}
\setlength\abovedisplayskip{18pt}
\setlength\belowdisplayskip{18pt}
V_{{\rm 2N}}^{(0)} & = & :\sum_{\bm{n}}\frac{1}{2}\Biggl(B_{1}\bar{\rho}^{2}+B_{2}\bar{\rho}_{i}\bar{\rho}_{i}\Biggr):,\nonumber \\
V_{{\rm 2N}}^{(2)} & = & :\sum_{\bm{n}}\frac{1}{2}\Biggl(-C_{1}\bar{\rho}\nabla^{2}\bar{\rho}-C_{2}\bar{\rho}_{a}\nabla^{2}\bar{\rho}_{a}-C_{3}\bar{\rho}_{i}\nabla^{2}\bar{\rho}_{i}\nonumber \\
 &  & -C_{4}\bar{\rho}_{ia}\nabla^{2}\bar{\rho}_{ia}-C_{5}\frac{1}{2}\epsilon_{ijk}(\bar{\rho}\partial_{i}\bar{\theta}_{k,j}+\bar{\rho}_{i}\partial_{j}\bar{\theta}_{0,k})\nonumber \\
 &  & -C_{6}\bar{\rho}_{i}\partial_{i}\partial_{j}\bar{\rho}_{j}-C_{7}\bar{\rho}_{ia}\partial_{i}\partial_{j}\bar{\rho}_{ja}\Biggr):,\nonumber \\
V_{1\pi} & = & :\sum_{\bm{n}}\frac{1}{2}\Biggl(C_{{\rm \pi}}\rho_{ia}\frac{\nabla_{i}\nabla_{j}\exp\left[\left(\nabla^{2}-M_{\pi}^{2}\right)/\Lambda_{\pi}^{2}\right]}{\nabla^{2}-M_{\pi}^{2}}\rho_{ja}\nonumber \\
 &  & +C_{\pi}^{\prime}C_{\pi}\rho_{ia}\exp\left[\left(\nabla^{2}-M_{\pi}^{2}\right)/\Lambda_{\pi}^{2}\right]\rho_{ia}\Biggr):,\label{eq:ops_spatial-1}
\end{eqnarray}
where repeated indices should be summed over. $C_{{\rm \pi}}=-g_{A}^{2}/\allowbreak(4F_{\pi}^{2})$
is the OPEP couping constant. The function of $\nabla$ could be understood
via the FFT.

\begin{eqnarray}
:e^{-a_{t}H}:& = & :e^{-a_{t}(K+V_{{\rm 2N}}^{(0)}+V_{{\rm 2N}}^{(2)}+V_{{\rm 1\pi}})}:\nonumber \\
 &=&\prod_{\bm{n}}\left(\int ds(\bm{n})\right):\exp\left\{ -a_{t}K\right.\nonumber\\
 & &\left.+\sum_{\bm{n}}\Biggl[-\frac{1}{2}\left(\alpha^2 + \sum_{i}(\beta_{i})^{2}+\gamma^{2}+\sum_{i}(\kappa_{i})^{2}\right.\right.\nonumber \\
 & &\left.+\sum_{i}(\lambda_{i})^{2}+\sum_{ia}(\mu_{ia})^{2}+|\zeta|^{2}+|\eta|^{2}+\xi^{2}\right.\nonumber\\
 & &\left.+\sum_{a}(\chi_{a})^{2}+\sum_{a}(\pi_{a})^{2}+\sum_{ia}(\delta_{ia})^{2}\right)\nonumber \\
 & &+\sqrt{-a_{t}B_{1}}\alpha\bar{\rho}+\sqrt{-a_{t}B_{2}}\beta_{i}\bar{\rho}_{i}\nonumber\\
 & &+\sqrt{-a_{t}C_{1}}\gamma\sqrt{-\nabla^{2}}\bar{\rho}+\sqrt{-a_{t}C_{2}}\kappa_{a}\sqrt{-\nabla^{2}}\bar{\rho}_{a}\nonumber \\
 & &+\sqrt{-a_{t}C_{3}}\lambda_{i}\sqrt{-\nabla^{2}}\bar{\rho}_{i}+\sqrt{-a_{t}C_{4}}\mu_{ia}\sqrt{-\nabla^{2}}\bar{\rho}_{ia}\nonumber\\
 & &-\frac{1}{2}\sqrt{-a_{t}C_{5}}\left(\zeta\bar{\rho}+\frac{1}{2}\zeta^{*}\epsilon_{ijk}\partial_{i}\bar{\theta}_{k,j}\right)\nonumber \\
 & &-\frac{1}{2}\sqrt{-a_{t}C_{5}}\left(\eta_{i}\bar{\rho}_{i}+\frac{1}{2}\eta_{i}^{*}\epsilon_{ijk}\partial_{j}\bar{\theta}_{0,k}\right)\nonumber\\
 & &+\sqrt{-a_{t}C_{6}}\xi\partial_{i}\bar{\rho}_{i}+\sqrt{-a_{t}C_{7}}\chi_{a}\partial_{i}\bar{\rho}_{ia}\nonumber \\
 & &+\sqrt{-a_{t}C_{\pi}}\pi_{a}\frac{\nabla_{i}\exp\left[(\nabla^{2}-M_{\pi}^{2})/(2\Lambda_{\pi}^{2})\right]}{\sqrt{-\nabla^{2}+M_{\pi}^{2}}}\rho_{ia}\nonumber\\
 & &+\sqrt{-a_{t}C_{\pi}^{\prime}C_{\pi}}\delta_{ia}\exp\left[\left(\nabla^{2}-M_{\pi}^{2}\right)/(2\Lambda_{\pi}^{2})\right]\rho_{ia}\Biggr]\Biggr\}:.\nonumber\\
 \label{eq:Auxiliary_field_expansion_of_H}
\end{eqnarray}

In Eq.~(\ref{eq:Auxiliary_field_expansion_of_H}) we present the auxiliary
field transformation of the interactions in Eq.~(\ref{eq:ops_spatial-1}).
Here $\alpha$, $\beta$, $\gamma$, $\kappa$, $\lambda$, $\mu$, $\xi$, $\chi$,
$\pi$ and $\delta$ are real auxiliary fields, $\zeta$ and $\eta$
are complex auxiliary fields. The indices $i$, $j$, $k$ and $a$
means independent components of the auxiliary fields and repeated
indices should be summed over. The integral over $s(\bm{n})$ means
integrating over all these continuous auxiliary fields for all lattice
sites. For complex auxiliary fields, we integrate out both the real and imaginary parts.

\subsection{Determination of the low-energy constants}

In this section, we determine the coefficients $B_{1-2}$, $C_{1-7}$,
$D_{1-15}$, $c_{{\rm nn, pp}}$ and $g_{1,2}$ by fitting
to the low-energy NN phase shifts and mixing angles.
The method is based on Ref.\cite{Phys.Lett.B760.309}. We decompose the scattering
waves on the lattice into different partial waves, then employ the
real and complex auxiliary potentials to extract the asymptotic radial
wave functions. 
We follow the conventional procedure for fitting the
LECs in the continuum\cite{PA747-362}. For a given order, we
first determine the operators to be included in the fitting procedures,
then find the spectroscopic LECs for each partial wave by a least
square fit against the Nijmegen phase shifts~\cite{PhysRevC.48.792} up to 200 MeV. The LECs
can be obtained by solving the linear equations. We also determine
the Galilean invariance restoration terms by requiring that the derivative
of the two $S$-wave scattering lengths against the center of mass
momentum squared $Q^{2}$ is exactly zero at $Q^{2}=0$. The charge
symmetry breaking LECs $c_{{\rm nn, pp}}$ are determined by fitting
the neuton-neutron and proton-proton scattering lengths to their empirical
values $a_{{\rm nn}}=-18.95$~fm and $a_{{\rm pp}}=-17.3$~fm, respectively.

\begin{figure*}[h]
\begin{centering}
\includegraphics[width=1.0\textwidth]{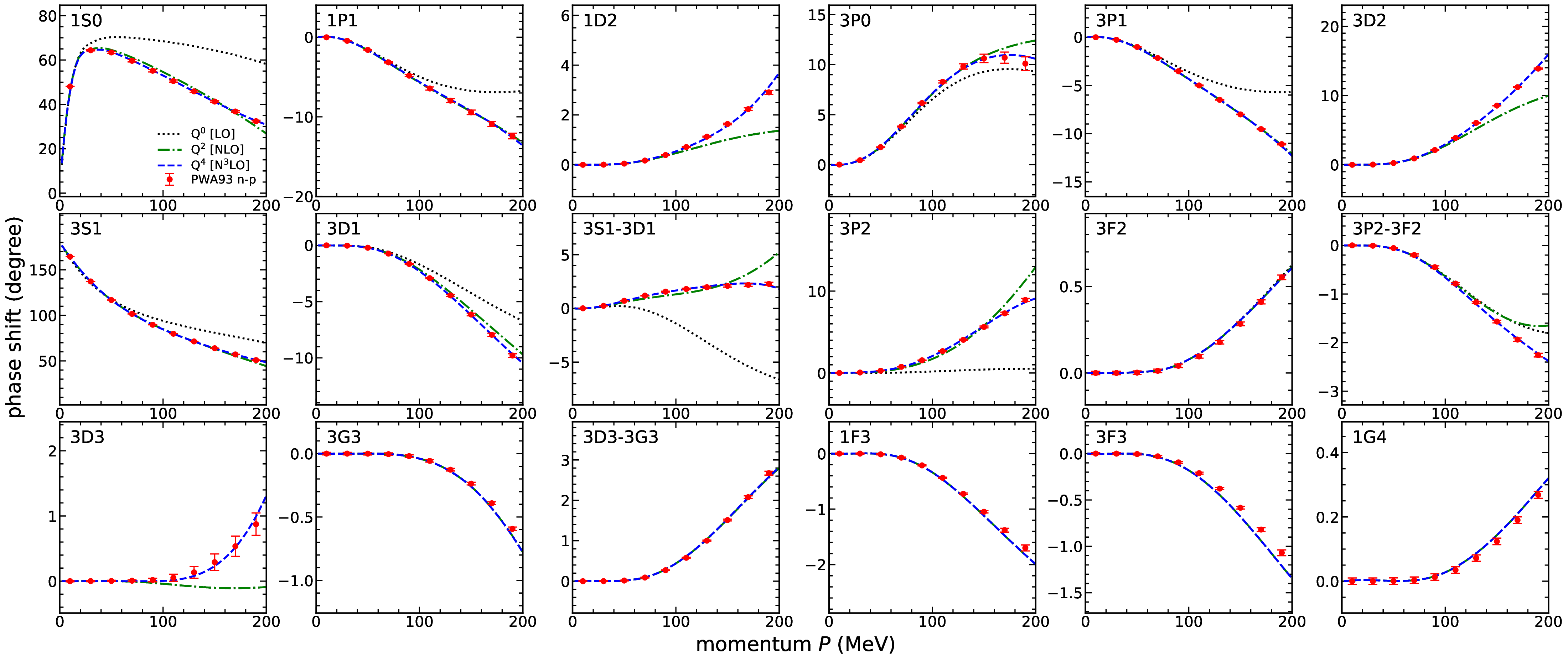}
\par\end{centering}
\caption{Calculated phase shifts for the LECs with $\Lambda = 400$ MeV in Table \ref{tab:Fitted-LECs-at}.
Dotted, dash-dotted and dashed lines denote the LO, NLO and N$^{3}$LO
results, respectively. 
Red circles with errorbars show the empirical
values~\cite{PhysRevC.48.792}. \label{fig:Calculated-phase-shifts}}
\end{figure*}
Following the conventional power counting scheme for chiral EFT, we add the charge symmetry breaking
terms and the Galilean invariance restoration terms to the next-to-leading order. Thus we have the following chain of chiral Hamiltonians with progressively
increasing precision,
\begin{eqnarray}
\setlength\abovedisplayskip{18pt}
\setlength\belowdisplayskip{18pt}
H_{{\rm LO}} & = & K+V_{{\rm 2N}}^{(0)},\nonumber \\
H_{{\rm NLO}} & = & K+V_{{\rm 2N}}^{(0)}+V_{{\rm 2N}}^{(2)}+V_{{\rm CSB}}^{{\rm nn}}+V_{{\rm CSB}}^{{\rm pp}}+V_{{\rm GIR}},\nonumber \\
H_{{\rm N3LO}} & = & K+V_{{\rm 2N}}^{(0)}+V_{{\rm 2N}}^{(2)}+V_{{\rm 2N}}^{(4)}+V_{{\rm CSB}}^{{\rm nn}}\nonumber \\
 &  & +V_{{\rm CSB}}^{{\rm pp}}+V_{{\rm GIR}},\label{eq:DefineHamiltonian}
\end{eqnarray}
where the LECs are different and should be fixed separately for Hamiltonians at different orders. 
As we have dropped the higher-order pion-exchange terms, there is no contribution at the N$^2$LO.
Note that the same operator can appear at different orders
with different coefficient. In Table \ref{tab:Fitted-LECs-at} we
show the resulting LECs for three different momentum cutoffs $\Lambda=300, 350$ and $400$ MeV. All results are in lattice unit. For comparison
with the continuum fits, the results should be multiplied with powers
of the lattice spacing $a$ to restore the dimension.
As an example, in Fig.~\ref{fig:Calculated-phase-shifts}
we show the calculated phase shifts with $\Lambda=400$ MeV compared with the empirical values.
The results for $\Lambda=300$ and 350~MeV are quite similar.
We find that the N$^{3}$LO interaction reproduces the phase shifts
with a high precision for the momentum interval 
considered here. 
In what follows we denote the ground state energies of the Hamiltonians in Eq.~(\ref{eq:DefineHamiltonian}) as $E_{\rm LO}$, $E_{\rm NLO}$ and $E_{\rm N3LO}$, respectively.
Similarly, the ground state energy of Eq.~(\ref{eq:HSU4}) is marked as $E_{\rm SU4}$.

\begin{table*}[h]
\caption{\label{tab:Fitted-LECs-at}Fitted LECs at each order with the momentum cutoff $\Lambda=300$, $350$ and $400$ MeV.
$B_{1-2}$, $C_{1-7}$ and $D_{1-15}$ denotes the LECs for the contact terms, $g_{1, 2}$ and $c_{\rm pp, nn}$ are LECs for Galilean invariance restoration terms and charge symmetry breaking terms, respectively.
The results are in the lattice unit system $\hbar=a=1$.}
\centering{}%
\resizebox{\linewidth}{!}{
\begin{tabular}{c|ccc|ccc|ccc}
\hline
\multicolumn{1}{c}{ }&\multicolumn{3}{c|}{$\Lambda=300$~MeV}&\multicolumn{3}{c|}{$\Lambda=350$~MeV}&\multicolumn{3}{c}{$\Lambda=400$~MeV}\\
\hline 
 & $H_{{\rm LO}}$ & $H_{{\rm NLO}}$ & $H_{{\rm N3LO}}$& $H_{{\rm LO}}$& $H_{{\rm NLO}}$&$H_{{\rm N3LO}}$& $H_{{\rm LO}}$ & $H_{{\rm NLO}}$ & $H_{{\rm N3LO}}$ \tabularnewline
\hline 
$B_{1}$ & $-3.6634$& $-4.5923$& $-5.1899$& $-3.0741$&$-4.2847$&$-4.3526$& $-2.6683$ & $-3.9971$ & $-3.4255$\tabularnewline
$B_{2}$ & $-0.2634$& $-0.2850$& $-0.3093$& $-0.1875$&$-0.2062$&$-0.0713$& $-0.1413$ & $-0.1188$ & $\quad0.1931$\tabularnewline
$C_{1}$ & & $\quad0.4544$& $\quad0.9130$& &$\quad 0.4493$&$\quad 0.7802$&  & $\quad 0.4315$ & $\quad 0.6576$\tabularnewline
$C_{2}$ & & $-0.0166$& $-0.1354$& & $-0.0415$& $-0.0482$&  & $-0.0555$ & $\quad0.0287$\tabularnewline
$C_{3}$ & & $-0.0334$& $-0.1699$& &$-0.0424$&$-0.1829$&  & $-0.0513$ & $-0.1916$\tabularnewline
$C_{4}$ & & $-0.0491$& $-0.1063$& &$-0.0672$&$-0.0650$&  & $-0.0780$ & $-0.0333$\tabularnewline
$C_{5}$ & & $\quad0.9012$& $\quad1.0230$& &$\quad0.8557$&$\quad1.0462$&  & $\quad0.8289$& $\quad1.0668$ \tabularnewline
$C_{6}$ & & $\quad0.0071$& $\quad0.1483$& &$-0.0019$&$\quad0.1555$&  & $-0.054$ & $\quad0.1545$\tabularnewline
$C_{7}$ & & $-0.2549$& $-0.4146$& &$-0.2450$&$-0.4361$&  & $-0.2468$ & $-0.4492$\tabularnewline
$D_{1}$ & & & $-0.1299$& & &$-0.1082$&  &  & $-0.0982$ \tabularnewline
$D_{2}$ & & & $\quad0.0232$& & &$\quad0.0081$&  &  & $\quad0.0009$\tabularnewline
$D_{3}$ & & & $\quad0.0290$& & &$\quad0.0245$&  &  & $\quad0.0232$\tabularnewline
$D_{4}$ & & & $\quad0.0075$& & &$-0.0016$&  &  & $-0.0064$\tabularnewline
$D_{5}$ & & & $-0.0266$& & &$-0.0258$&  &  & $-0.0260$\tabularnewline
$D_{6}$ & & & $\quad0.0504$& & &$\quad0.0524$& &  & $\quad0.0518$\tabularnewline
$D_{7}$ & & & $-0.1920$& & &$\quad0.0171$&  &  & $\quad0.0726$\tabularnewline
$D_{8}$ & & & $-0.3799$& & &$-0.3845$&   &  & $-0.3787$\tabularnewline
$D_{9}$ &  & & $\quad1.1896$& & &$\quad0.6709$&  &  & $\quad0.4582$\tabularnewline
$D_{10}$ & &  & $\quad0.1549$& & &$\quad0.1090$&  &  & $\quad0.0825$\tabularnewline
$D_{11}$ &  & & $-0.0663$& & &$-0.0939$&   &  & $-0.1076$\tabularnewline
$D_{12}$ & & & $-0.0295$& & &$-0.0348$&   &  & $-0.0259$\tabularnewline
$D_{13}$ & & & $\quad0.2975$& & & $\quad0.3236$&   &  & $\quad0.3228$\tabularnewline
$D_{14}$ &  & & $\quad0.4141$& & &$\quad0.4091$&  &  & $\quad0.4030$\tabularnewline
$D_{15}$ & & & $\quad0.3805$& &&$\quad0.4721$&   &  & $\quad0.4820$\tabularnewline
$g_{1}$ & & $-0.5005$& $-0.4122$& &$-0.1929$&$-0.1595$&  & $-0.0782$ & $-0.0696$\tabularnewline
$g_{2}$ & & $-0.0822$& $-0.0765$& &$-0.0267$&$-0.0225$&  & $-0.0051$ & $\quad0.0077$\tabularnewline
$c_{\text{pp}}$ & &$\quad0.1425$&$\quad0.1504$& &$\quad0.1271$&$\quad0.1398$& &$\quad0.1241$&$\quad0.1450$\tabularnewline
$c_{\text{nn}}$ & &$\quad0.0633$&$\quad0.0717$& &$\quad0.0578$&$\quad0.0670$& &$\quad0.0567$&$\quad0.0690$\tabularnewline
\hline 
\end{tabular}}
\end{table*}

\subsection{Perturbative quantum monte carlo method\label{Sec:ptQMC}}

In this section, we briefly introduce the perturbative quantum Monte
Carlo (ptQMC) method and its application for solving the chiral Hamiltonian.
The details of the ptQMC method for N$^2$LO Hamiltonian has been given in Ref.~\cite{ref11_in_PRL128-242501} and here we focus on its extension for the N$^3$LO Hamiltonian.
In NLEFT the ground state of any nucleus can be calculated using the
imaginary time projection method,
\begin{equation}
\setlength\abovedisplayskip{18pt}
\setlength\belowdisplayskip{18pt}
|\Psi\rangle\propto\lim_{\tau\rightarrow\infty}\exp(-H\tau)|\Psi_{T}\rangle,\label{eq:gswavefunction}
\end{equation}
where $|\Psi_{T}\rangle$ is an arbitrarily chosen trial wave function
that has non-zero overlap with the real ground state, $H$ is the
full Hamiltonian. The expectation value of any observable can be calculated
accordingly,
\[
\langle O\rangle=\lim_{\tau\rightarrow\infty}\frac{\langle\Psi_{T}|\exp(-H\tau/2)O\exp(-H\tau/2)|\Psi_{T}\rangle}{\langle\Psi_{T}|\exp(-H\tau)|\Psi_{T}\rangle},
\]
where the imaginary time projection operators can be applied by discretizing and 
extrapolating the variable $\tau$. Then, we decompose the interactions in $H$ using the auxiliary field transformation, rewriting the nucleon-nucleon couplings as interactions between the nucleons and external auxiliary fields.
The nuclear forces can be recovered by integrating out the auxiliary fields at every lattice sites. 
For a simple two-body contact term the transformation reads
\begin{equation}
:e^{-\tfrac{1}{2}a_{t}C_{0}\rho({\bm{n}})^{2}}: \propto\int\mathcal{D}s:e^{-\tfrac{s(\bm{n})^{2}}{2}+\sqrt{-a_{t}C_{0}}s(\bm{n})\rho(\bm{n})}
\label{eq:HS_transform-1}
\end{equation}
where $a_{t}$ is a tiny temporal step, $C_{0}$ is the coupling constant,
$s(\bm{n})$ is a scalar auxiliary field, $\rho(\bm{n})=\hat{a}(\bm{n})^{\dagger}\hat{a}(\bm{n})$
is the density operator for the nucleons and the summation over $\bm{n}$
runs through all lattice sites.
In this work we take $a_t=(1000$~MeV$)^{-1}$ throughout and denote the total number of time slices as $L_t$. 
Applying Eq.~(\ref{eq:HS_transform-1})
repeatedly on $|\Psi_{T}\rangle$ by $L_t/2$ times we end up with a path integral,
\begin{equation}
\setlength\abovedisplayskip{18pt}
\setlength\belowdisplayskip{18pt}
\langle O\rangle=\lim_{\tau\rightarrow\infty}\frac{\int\mathcal{D}s\langle\Psi_{T}|\cdots O \cdots|\Psi_{T}\rangle}{\int\mathcal{D}s\langle\Psi_{T}|\cdots |\Psi_{T}\rangle},\label{eq:pathintegral}
\end{equation}
where the elipsis denotes the integrand in Eq.~(\ref{eq:HS_transform-1}).
For different temporal steps the $s$ fields are independent of each
other. The ratio of the path integral in Eq.~(\ref{eq:pathintegral}) can be evaluated
using the Monte Carlo algorithm with the importance sampling.

To apply the ptQMC method, we assume that $H_{0}$ has already captured all the essential elements
of $H$ and the residual $V_{C}=H-H_{0}$ can be treated as perturbations.
Now we expand the real ground state wave function  Eq.~(\ref{eq:gswavefunction}) of the full Hamiltonain $H$
into a Taylor series up to the linear term of the residual Hamiltonian, 
\begin{align*}
\setlength\abovedisplayskip{18pt}
\setlength\belowdisplayskip{18pt}
|\Psi\rangle & =\lim_{L_{t}\rightarrow\infty}M^{L_{t}/2}|\Psi_{ T}\rangle=|\Psi_{0}\rangle+|\Psi_{1}\rangle+\mathcal{O}(V_{C}^{2}),
\end{align*}
where $M=:\exp(-a_t H):$ is the full transfer matrix, the numbers in the subscripts denote the orders in the perturbation
theory,
\begin{align*}
\setlength\abovedisplayskip{18pt}
\setlength\belowdisplayskip{18pt}
|\Psi_{0}\rangle & =\lim_{L_{t}\rightarrow\infty}M_{0}^{L_{t}/2}|\Psi_{T}\rangle,\\
|\Psi_{1}\rangle & =\lim_{L_{t}\rightarrow\infty}\frac{2}{L_{t}}\sum_{k=1}^{L_{t}/2}M_{0}^{L_{t}/2-k}(M-M_{0})M_{0}^{k-1}|\Psi_{T}\rangle,
\end{align*}
where we have used the identity 
\begin{equation}
M=:M_{0}(1-a_{t}V_{C}):+\mathcal{O}(V_{C}^{2}).
\end{equation}
The normalized ground state wave function up to $\mathcal{O}(V_{C})$
is
\begin{align}
\setlength\abovedisplayskip{18pt}
\setlength\belowdisplayskip{18pt}
|\overline{\Psi}\rangle & =\frac{|\Psi\rangle}{\sqrt{\langle\Psi|\Psi\rangle}}=\frac{|\Psi_{0}\rangle}{\sqrt{\langle\Psi_{0}|\Psi_{0}\rangle}}+\frac{1}{\sqrt{\langle\Psi_{0}|\Psi_{0}\rangle}}\nonumber \\
 & \times\left[|\Psi_{1}\rangle-\frac{{\rm Re}\langle\Psi_{1}|\Psi_{0}\rangle}{\langle\Psi_{0}|\Psi_{0}\rangle}|\Psi_{0}\rangle\right]+\mathcal{O}(V_{C}^{2}),\label{eq:norm_wave-1-1-1}
\end{align}
where ${\rm Re}$ means taking the real part. Eq.~(\ref{eq:norm_wave-1-1-1})
can be used to calculate the expectation value of any operator up
to $\mathcal{O}(V_{C})$. For a general observable $O$ we have
\begin{equation}
\setlength\belowdisplayskip{18pt}
\langle O\rangle = O_{0}+2{\rm Re}\left(\frac{\langle\Psi_{0}|O|\Psi_{1}\rangle} {\langle \Psi_0|\Psi_0\rangle} -O_{0}\frac{\langle\Psi_{0}|\Psi_{1}\rangle} {\langle \Psi_0|\Psi_0\rangle} \right)  +\mathcal{O}(V_{C}^{2}) \label{eq:Oexpansion}
\end{equation}
where $O_{0}=\langle\Psi_{0}|O|\Psi_{0}\rangle/\langle\Psi_{0}|\Psi_{0}\rangle$
is the zeroth order expectation value. A special case is the energy,
for which the first order correction only depends on the
zeroth order wave function $|\Psi_{0}\rangle$ and we can compute
the second order perturbative energy using $|\overline{\Psi}\rangle$,
as was demonstrated in Ref.\cite{PhysRevLett.128.242501}. 
This is because the $H$
inserted in the middle time step also contains a term linear in $V_{C}$, which modifies the power counting of the perturbative senries over
$V_{C}$.

As was shown in Ref.\cite{PhysRevLett.128.242501}, the matrix elements in Eq.~(\ref{eq:Oexpansion}) containing
$|\Psi_{1}\rangle$ can be computed efficiently using the auxiliary
field transformation of the interactions in $V_{C}$. The resulting
auxiliary fields can be complexified and shifted to reduce the sign
problem, then integrated out with an importance sampling method. Note
that for applying the ptQMC method we only require that $V_{C}$ can
be decomposed using the auxiliary fields, whether these auxiliary
fields cause sign problem in a non-perturbative calculation is not an
essential issue.
More specifically, for every sample $\{s_{1},s_{2},\cdots,s_{L_{t}}\}$ with $s_{k}$ the auxiliary fields in Eq.~(\ref{eq:HS_transform-1}), 
we have
\begin{equation}
\langle \Psi_0 | O |\Psi_1\rangle=\int\mathcal{D}cP(c+\bar{c})\langle\cdots O\cdots M(s_{k},c+\bar{c})\cdots\rangle_{T},\label{eq:shifted_integral}
\end{equation}
where the ellipses denote the transfer matrices $M_{0}(s_{t})$
with $t\neq k$, $\langle\rangle_{T}$  the expectation value in the
state $|\Psi_{T}\rangle$, $c$ the auxiliary fields for interactions  in $V_C$ and $P(c)$ is the standard normal distribution.
In Eq.~(\ref{eq:shifted_integral}) we have made a variable change $c\rightarrow \bar{c} + c$
with $c$ real integral variables.
Here $\bar{c}(\bm{n})$ is a constant complex-valued field
\begin{align}
\bar{c}(\bm{n}) & =\left.\frac{\partial}{\partial c(\bm{n})}\ln\langle\cdots M(s_{k},c)\cdots\rangle_{T}\right|_{c=0}\nonumber \\
 & =\sqrt{-a_{t}C}\langle\cdots:M_{0}(s_{k})\rho_{c}(\bm{n}):\cdots\rangle_{T}/\langle \cdots \rangle_T \label{eq:stationary_point_condition}
\end{align}
where the ellipses represent the zeroth-order transfer matrices $M_{0}$'s, $C$ is the coupling
constant for the $V_{C}$ term and $\rho_c$ represents the corresponding density. Generally, $\bar{c}$ is a complex
field, \textit{e.g.}, for repulsive interactions such as Coulomb we have $C>0$,
the square root in Eq.~(\ref{eq:stationary_point_condition}) introduces
an imaginary factor $i$, which causes severe sign problem in non-perturbative calculations.
We can
use stochastic methods to evaluate Eq.~(\ref{eq:shifted_integral})
by sampling the $c$ field with a standard normal distribution. 
The
variable change in Eq.~(\ref{eq:shifted_integral}) can reduce the
statistical error by one order or more.
See Ref.~\cite{PhysRevLett.128.242501} for details of the proof and the numerical demonstration.

We now consider a general Hamiltonian consisting of many terms such
as $H_{{\rm N3LO}}$ presented in Eq.~(\ref{eq:DefineHamiltonian}). There are two-fold difficulties in solving it
with the NLEFT. Firstly, there may be interactions causing severe sign
problem, preventing a full non-perturbative calculation. This can
be partly solved by searching for a non-perturbative Hamiltonian without
the sign problem such as $H_{{\rm SU4}}$ and calculating the energy
corrections using the ptQMC method. Secondly, find the auxiliary field
transformation for each term in the Hamiltonian and the corresponding
complexification may be unfeasible or too laborious. For example,
a full decomposition of $H_{{\rm N3LO}}$ requires more than $70$
auxiliary fields, containing real, complex, discrete and angle variables.
To simplify the calculation, we propose the following scheme for a
general ptQMC calculation with a hierarchical nuclear force,
\begin{enumerate}
\item Find a sign-problem-free Hamiltonian $H_{0}$ close to $H$. Solve
$H_{0}$ using the Monte Carlo method. The resulting energy is the
zeroth order contribution $E_{0}$.
\item Find an improved Hamiltonian $H_{1}$ satisfying two conditions: 1)
$H_{1}$ is a better approximation of $H$ than $H_{0}$; 2) The interactions
in $H_{1}$ can be efficiently decomposed using the auxiliary field
transformation. Then the difference $V_{C}=H_{1}-H_{0}$ can be included
up to the second order $\mathcal{O}(V_{C}^{2})$ using the standard
ptQMC method.
\item The remaining difference $V_{C}^{\prime}=H-H_{1}$ can be taken into
account by inserting it into Eq.~(\ref{eq:Oexpansion}) with $V_{C}$
defined in the previous step. Note that $V_{C}^{\prime}$ is only
inserted in the middle time step, its
auxiliary field transformation is not needed. In this step we find the corrections
up to $\mathcal{O}(V_{C}V_{C}^{\prime})$.
\end{enumerate}

\begin{figure}
\begin{centering}
\includegraphics[width=1\columnwidth]{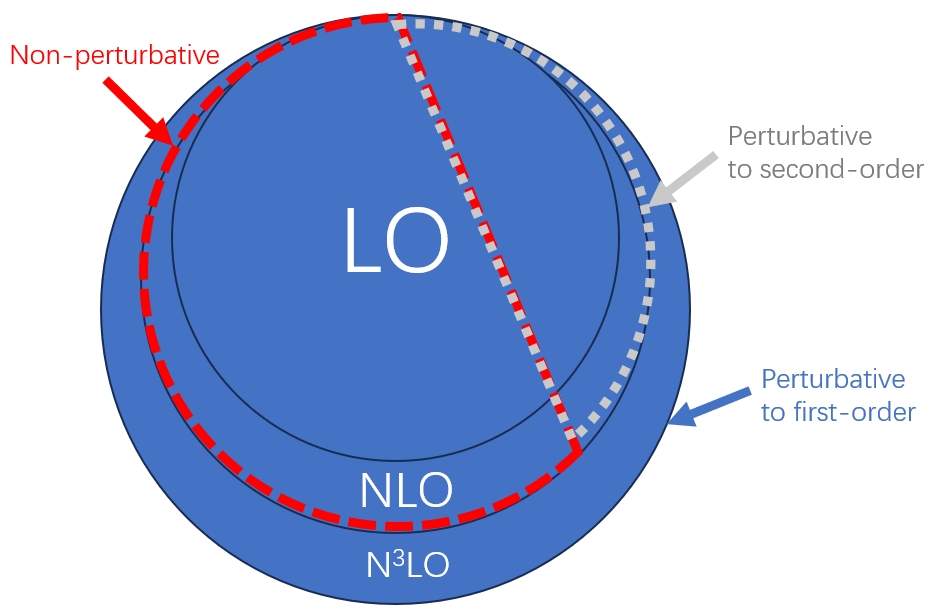}
\par\end{centering}
\caption{Schematic plot for the ptQMC calculation of the N$^3$LO Hamiltonian. 
We split the Hamiltonian into various pieces and handle them with different orders of perturbation theory to balance the precision and efficiency of the calculation. 
\label{fig:ptQMC_schematic}}
\end{figure}

Compared with a standard ptQMC calculation, we immediately find that
we have dropped the term $\mathcal{O}(V_{C}^{\prime2})$. This
operation can be justified only if $V_{C}^{\prime}$ is much smaller
than $V_{C}$. This is just the case for the chiral forces. We have
already known that $H_{{\rm SU4}}$ captures the most essential part
of the nuclear force and $H_{{\rm NLO}}$ accounts for the specific
spin and isospin dependencies, we can set $H_{0}=H_{{\rm SU4}}$,
$H_{1}=H_{{\rm NLO}}$ and apply the above protocols to solve the
more complicated $H_{{\rm N3LO}}$. We expect that the residual $H_{{\rm N3LO}}-H_{{\rm NLO}}$
is so small that only its linear contributions are important. 
This assumption corresponds to a double expansion with the expression
\begin{equation}
H_{{\rm N3LO}}=H_{{\rm SU4}}+\lambda(H_{{\rm NLO}}-H_{{\rm SU4}})+\mu(H_{{\rm N3LO}}-H_{{\rm NLO}}), \label{eq:N3LOexpansion}
\end{equation}
where $\lambda$ and $\mu$ are small parameters inserted for convenience.
For $\lambda=\mu=1$, we recover the full N$^3$LO Hamiltonian.
As is shown in Eq.~(\ref{eq:Auxiliary_field_expansion_of_H}), we can rewrite the interactions in $H_{\rm NLO}$ using the auxiliary fields and apply the ptQMC method to find the perturbative energy at the $\mathcal{O}(\lambda^2)$ order.
For the term proportional to $\mu$ we simply apply Eq.~(\ref{eq:Oexpansion}) to calculate the perturbative energy at the $\mathcal{O}(\lambda\mu)$ order.
Note that in Eq.~(\ref{eq:Oexpansion}) the wave function $|\Psi_1\rangle$ is proportional to $\lambda$ and $\mathcal{O}(\mu^2)$ term is not considered.
In what follows, we denote the ground state energy of Eq.~(\ref{eq:N3LOexpansion}) as $E_{\rm N3LO}(\lambda,\mu)$.
The perturbative calculation means a Taylor expansion
\begin{equation}
    E_{\rm N3LO}(\lambda,\mu) = \sum_{n,m=0}^{\infty} E_{\lambda^n\mu^m} \lambda^n \mu^m.
\end{equation}
In numerical benchmark we also consider the expansion of $E_{\rm NLO}$ around $E_{\rm SU4}$ with a single parameter $\lambda$.
The symbols like $E_{\lambda^n}$ are defined similarly.

In Fig.~\ref{fig:ptQMC_schematic} we schematically plot the decomposition of the full N$^3$LO Hamiltonian.
The three nested circles represent the Hamiltonian $H_{\rm LO}$, $H_{\rm NLO}$ and $H_{\rm N3LO}$, respectively.
The area enclosed by the red dashed lines denotes the Hamiltonian $H_{\rm SU4}$ to be solved non-perturbatively.
Note that the parameters in the non-perturbative part can be optimized to minimize the residual part $H_{\rm NLO} - H_{\rm SU4}$.
The grey dotted line encloses $H_{\rm NLO} - H_{\rm SU4}$ which is calculated up to the second order.
Here we can make a simple correspondence to understand which part of $H_{\rm NLO}$ has been absorbed into $H_{\rm SU4}$.
The local smearing Eq.~(\ref{eq:localsmearing}) acts on the density operators and mimics the effect of the $q^2$ term at NLO, while the non-local smearing Eq.~(\ref{eq:nonlocalsmearing}) acts on the single-particle degrees of freedom and mimics the effect of the $k^2$ term at NLO.
The other spin and isospin dependent terms are not absorbed into $H_{\rm SU4}$.
Note that such a mapping is not exact.
The smearing operations also introduce additional SU4 operators such as the Galilean invariance breaking terms.
The remaining piece beyond $H_{\rm NLO}$ in Fig.~\ref{fig:ptQMC_schematic} corresponds to $H_{\rm N3LO} - H_{\rm NLO}$ to be included up to the first order.
The general rule is that the wave functions are always corrected with the NLO Hamiltonian and all other corrections are inserted into the middle time step as first order corrections to the energy.

\section{Results and Discussions}

\begin{figure}
\begin{centering}
\vspace{-1em}
\includegraphics[width=1\columnwidth]{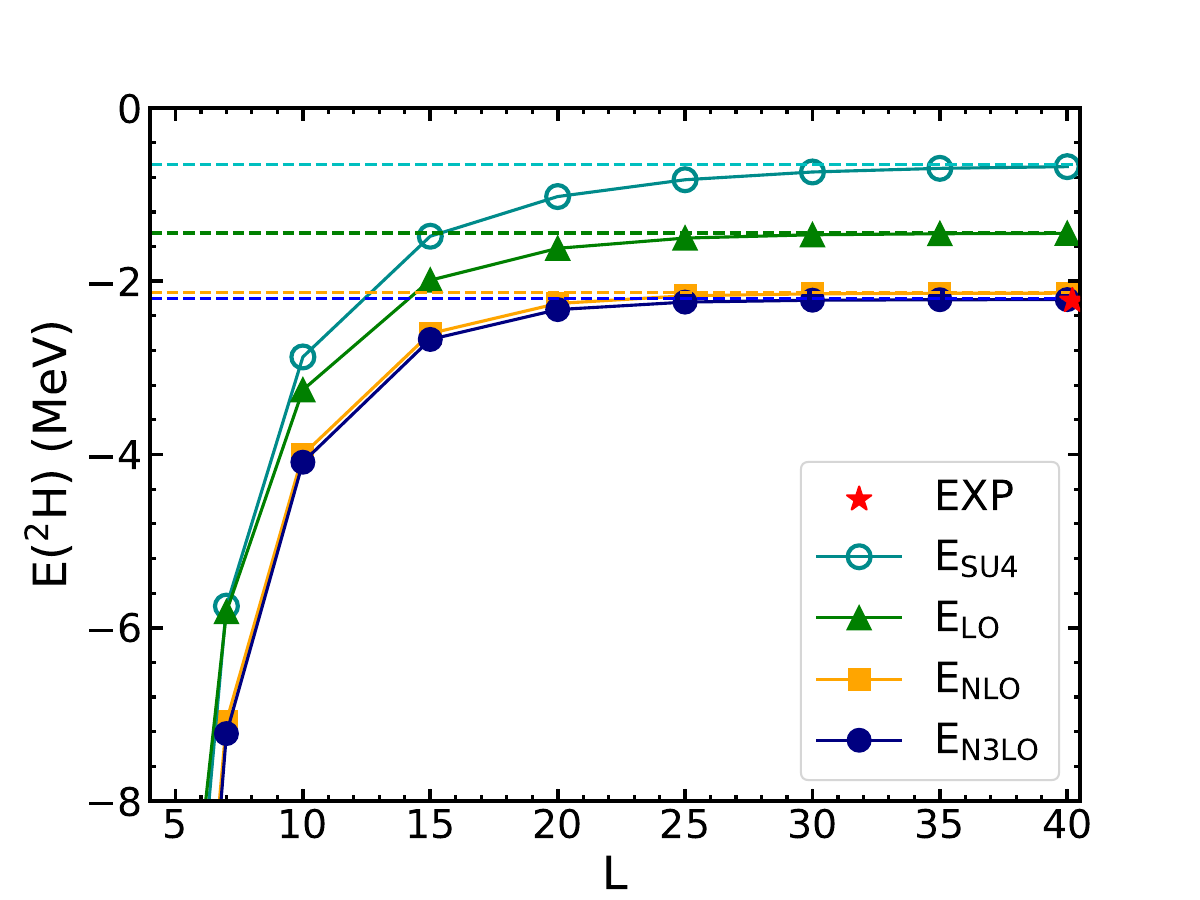}
\par\end{centering}
\caption{The ground state energies of the deuteron as functions of the box size $L$.
The full symbols denote the results calculated with the Hamiltonians defined in Eq.~(\ref{eq:DefineHamiltonian}) with the momentum cutoff $\Lambda=$ 400 MeV.
The open circles display the results from the SU4 Hamiltonian Eq.~(\ref{eq:HSU4}).
The extrapolated results for $L\rightarrow\infty$ are denoted by the dashed lines.
The experimental value is marked as a red star.
\label{fig:groundstateH2}}
\end{figure}

In this section we benchmark the algorithm. We start with the deuteron,
for which the energies can be solved exactly with the Lanczos algorithm
and the perturbative results can be compared with the exact values
order by order. In Fig.~\ref{fig:groundstateH2} we show the calculated
ground state energy of $^{2}$H using different chiral Hamiltonian
defined in Eq.~(\ref{eq:DefineHamiltonian}) as well as the SU4 Hamiltonian given in Eq.~(\ref{eq:HSU4}). Note that for two-body
neutron-proton system in the center of mass frame, 
the Coulomb force $V_{{\rm cou}}$, the Galilean invariance restoration
term $V_{{\rm GIR}}$ and charge symmetry breaking term $V_{{\rm CSB}}$ vanish completely. The experimental deuteron binding
energy $E(^{2}$H$)=-$2.22 MeV is so small that the wave function has a large spatial extension
and requires a large volume to suppress the finite volume effects.
Here we examine the dependencies of the results on the box size and
make extrapolations to the infinite volume limit $L\rightarrow\infty$.
The extrapolated energies are $-$0.66 MeV, $-$1.44 MeV, $-2.13$
MeV and $-2.19$ MeV for $H_{{\rm SU4}}$, $H_{{\rm LO}}$, $H_{{\rm NLO}}$
and $H_{{\rm N3LO}}$, respectively. We can see a clear convergence
pattern towards the experimental value. As we have not included the
deuteron binding energy as an additional constraint in the fitting
procedure, this quantity is a pure prediction and reflects the quality
of our phaseshift fits. 

Unfortunately, the physical deuteron is not a suitable test ground for perturbation theory. As its binding energy is close to threshold
between the bound states and the continuum, the perturbative series
converge slowly or even diverge due to the existence of a branching
point corresponding to the critical value of the coupling strength.
To avoid such non-essential complexities, here we enclose the deuteron
in a small volume $L=7$ fm with periodic boundary conditions. In such
a small box the deuteron binding energies are $-$5.75 MeV, $-$5.81
MeV, $-$7.08 MeV and $-$7.22 MeV for $H_{{\rm SU4}}$, $H_{{\rm LO}}$,
$H_{{\rm NLO}}$ and $H_{{\rm N3LO}}$, respectively. In this case
the ground state is well isolated from the continuum and other eigenstates
of the Hamiltonian, thus the perturbative corrections are suppressed
by powers of $1/|E_{{\rm g.s.}}-E_{{\rm ex.}}|$ and converge quickly,
where $E_{{\rm g.s.}}$ and $E_{{\rm ex.}}$ are energies of the ground
state and first excited state, respectively. This estimation also
applies to general perturbative calculations in many-body systems. We expect that the
ptQMC method exhibits the highest precision for deeply bound nuclei
such as $^{4}$He and $^{16}$O, but becomes less efficient for shallow
bound states such as physical deuteron and $^{6}$He. For the latter
case the ptQMC method must be improved to take into account the non-analytical
structure of the energies as functions of the coupling constants.

\subsection{Convergence pattern of the perturbative series\label{Sec:HNLOexpansion}}

\begin{figure}
\begin{centering}
\includegraphics[width=1\columnwidth]{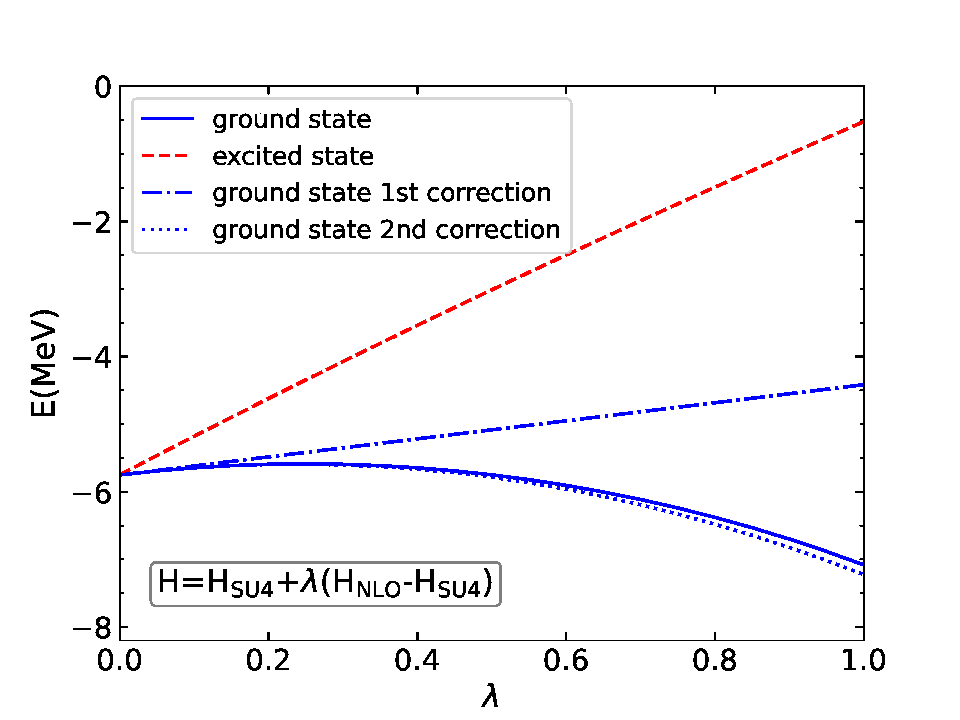}
\par\end{centering}\vspace{-0.5em}
\caption{The binding energies of the ground (solid line) and the first excited (dashed line) states of the deuteron enclosed in a $L=7$ box as functions of the small parameter $\lambda$ defined in Eq.~(\ref{eq:Hlambda}).
The dots denote the values of $\lambda$ at which we calculate the energies.
The dashed-dotted line and dotted lines represent the first and second order perturbative approximations of the ground state energies.
\label{fig:H2states}}
\end{figure}

We first consider the solution of the chiral Hamiltonian $H_{{\rm NLO}}$
by expanding it around $H_{{\rm SU4}}$. In Fig.~\ref{fig:H2states}
we plot the first two energy levels of the interpolated Hamiltonian
\begin{equation}
H(\lambda)=H_{{\rm SU4}}+\lambda(H_{{\rm NLO}}-H_{{\rm SU4}}),  \label{eq:Hlambda}
\end{equation}
where $0\leq\lambda\leq1$ is a real parameter. For $\lambda=0$ we
start with $H_{{\rm SU4}}$ and for $\lambda=1$ we recover the NLO
chiral Hamiltonian. The perturbative calculation corresponds to a
Taylor expansion of $\lambda$ around $\lambda=0$. In infinite volume
the chiral Hamiltonian $H_{{\rm NLO}}$ commutes with the total spin
squared $\hat{S}^{2}$ and the ground state is just the deuteron with
the corresponding quantum number $S=1$. This observation also holds
for deuteron in a finite volume. By diagonalizing $H_{{\rm NLO}}$
at $L=7$ we find three degenerate ground states and one non-degenerate excited states slightly
below the threshold $E=0$. By applying the operator $\hat{S}^{2}$
we found that these two energy levels correspond to the $^{3}$S$_{1}$
and $^{1}$S$_{0}$ channels in the infinite volume limit, respectively.
For the spin-independent Hamiltonian $H_{{\rm SU4}}$ these two channels
feel exactly the same interactions and the lowest four energy levels
are completely degenerate, which is reflected by the convergence of
the lines at $\lambda=0$ in Fig.~\ref{fig:H2states}. Consequently,
in principle we need to employ the degenerate perturbation theory
. Alternatively, for the ground state calculations,
we can either manually exclude the state with the wrong quantum numbers
or add a correction term like $k(1-\bm{\sigma}_{1}\cdot\bm{\sigma}_{2})$
with $k$ a large positive number to push the $^{1}$S$_{0}$ states
to very high energies. For projection Monte Carlo calculations, we can 
build the trial wave functions with two nucleon spins pointing to
the same direction, then only the spin-triplet states can be projected.
 The perturbative ground state energies are depicted
in Fig.~\ref{fig:H2states} with blue dash-dotted straight line and
blue dotted parabola for first and second order perturbative approximations,
respectively. 
The agreement between second-order perturbative energies and exact values implies that any potential improvements from third- or higher-order corrections would likely be marginal.

It is interesting to examine the convergence of the perturbative series
beyond the second order. As we can find the exact energies for any
value of $\lambda$ by direct diagonalization, the energy corrections
at each order can be obtained by evaluating the third and higher order
derivatives at $\lambda=0$ numerically. However, the finite difference
formulae for higher-order derivatives becomes too complicated and
the resulting numerical error increases exponentially against the
perturbative order. Here we use the Cauchy integral to calculate the
derivatives instead,
\begin{equation}
\setlength\abovedisplayskip{18pt}
\setlength\belowdisplayskip{18pt}
\left.\frac{\partial^{n}E}{\partial\lambda^{n}}\right|_{\lambda=0}=\frac{n!}{2\pi i}\oint_{|\lambda|=1}\frac{E(\lambda)}{\lambda^{n+1}}d\lambda,\label{eq:CauchyFormula}
\end{equation}
where the integral is performed around the unit circle on the complex-$\lambda$ plane.
\begin{figure}
\begin{figure}[H]
    \centering
    \begin{minipage}{0.9\columnwidth}
        \includegraphics[width=\linewidth]{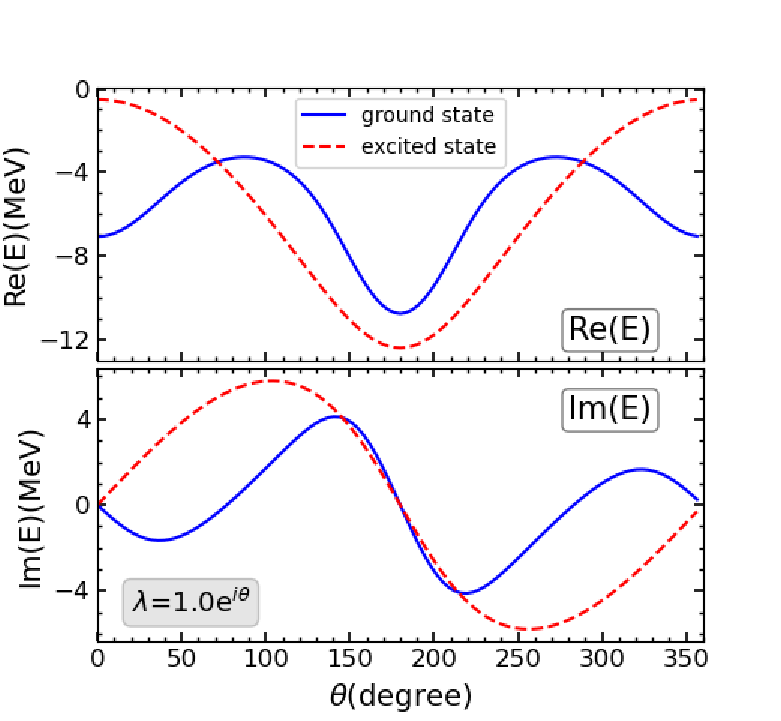}
    \end{minipage}
\end{figure}\vspace{-0.9em}
\caption{The ground (solid lines) and first excited (dashed lines) state energies of the deuteron enclosed in a $L=7$ box as functions of the argument $\theta=$arg$(\lambda)$.
The small parameter $\lambda$ is defined in Eq.~(\ref{eq:Hlambda}) and circulated around the unit circle $|\lambda|=1$ on the complex $\lambda$-plane.
The upper and lower panels denote the real and imaginary parts of the energies, respectively.
\label{fig:complexLambdaPlane}}
\end{figure}
We choose $20$ points uniformly distributed
on the unit circle $|\lambda|=1$ and the integral in Eq.~(\ref{eq:CauchyFormula})
becomes a summation over the corresponding energies $E(\lambda)$
multiplied by a Fourier transform factor $e^{-i(n+1)\theta}$. Note
that for a complex $\lambda$ the Hamiltonian $H(\lambda)$ is no
longer Hermitian and the eigenvalues are also complex. In  Fig.~\ref{fig:complexLambdaPlane} we show the real and imaginary
parts of $E(\lambda)$ as functions of the argument $\theta=\arg(\lambda)$.
For ground state the function $E(\lambda)$ is a smooth periodic function
of $\theta$, whose contour integral can be precisely calculated
with the uniform sampling points and then applied to extract the derivatives
Eq.~(\ref{eq:CauchyFormula}). Note that for such calculations the
energy level $E(\lambda)$ to be integrated must be chosen carefully
as the ordering of the low-lying energy levels
can change significantly for complex $\lambda$. 
To illutraste this point we also plot the
first excited state as dashed lines in the right panel of Fig.~\ref{fig:complexLambdaPlane}.
We see that for $\theta\approx200^{\circ}$ the excited state instead
of the ground state has the lowest real part of the energy. In this
work we use the fact that $E(\lambda)$ is a continous function of
$\theta$ to identify the energy level to be included. We start from
$\theta=0$, for each value of $\theta$ we compare the eigen functions
with the previous wave function with smaller $\theta$,
then choose the one with the largest overlap.

\begin{figure}
\begin{centering}
\includegraphics[width=1.02\columnwidth]{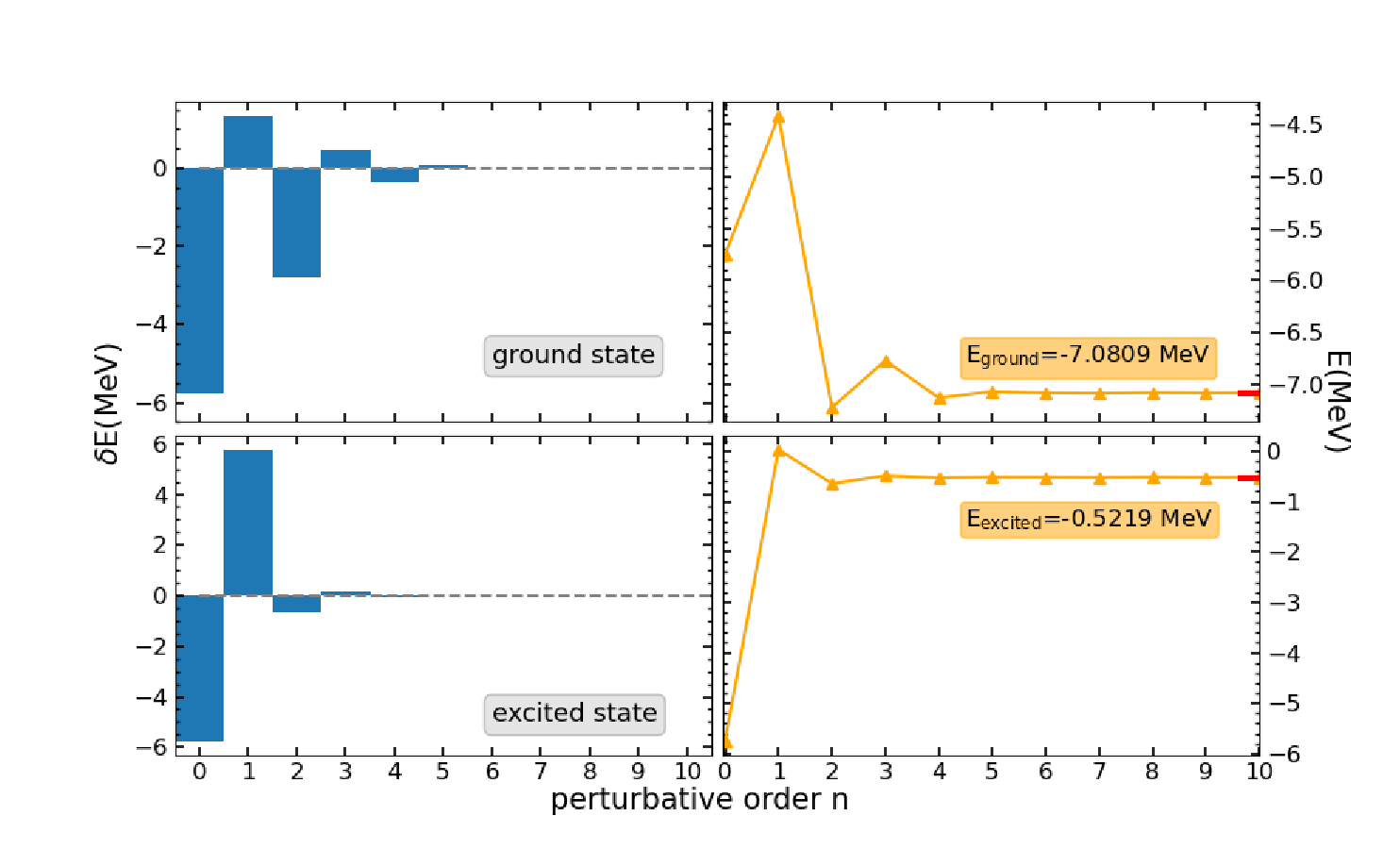}
\par\end{centering}
\caption{Left panels: The perturbative corrections at each order for the ground (upper panel) and first excited (lower panel) state energies of the deuteron enclosed in a $L=7$ box.
The zeroth order Hamiltonian is $H_{\rm SU4}$ and 
the target Hamiltonian is $H_{\rm NLO}$ with the momentum cutoff $\Lambda=400$ MeV.
Right panels: The partial sums of the perturbative corrections up to each orders. 
The numbers in the boxes represent the converged values.
\label{fig:DeuteronPartialCOntribution}}
\end{figure}

In the left panels of Fig.~\ref{fig:DeuteronPartialCOntribution} we
show the contributions to the total binding energies from each perturbative
orders. In the right panel we show the partial sums up to the corresponding
orders. As we have already obtained $E(\lambda)$ for both the ground
state and first excited state in Fig.~\ref{fig:complexLambdaPlane},
here we calculate and show the perturbative expansion for both states.
We see a clear convergence pattern in both cases. For ground state
$H_{{\rm SU4}}$ gives a dominant zeroth order energy, while the first
and second order corrections are also significant. The contributions
from the third and fourth orders are relatively unimportant and cancel
each other. The contributions from $n\geq6$ are essentially zero.
The sum of these corrections converge quickly to the exact values.
For the excited state the zeroth and first order energies have opposite
signs and mostly cancel each other, while the second and higher order
corrections are much smaller. This is consistent with the results
in Fig.~\ref{fig:H2states}, where the red dashed line is approximately
straight.

\begin{figure}
\begin{centering}
\includegraphics[width=1\columnwidth]{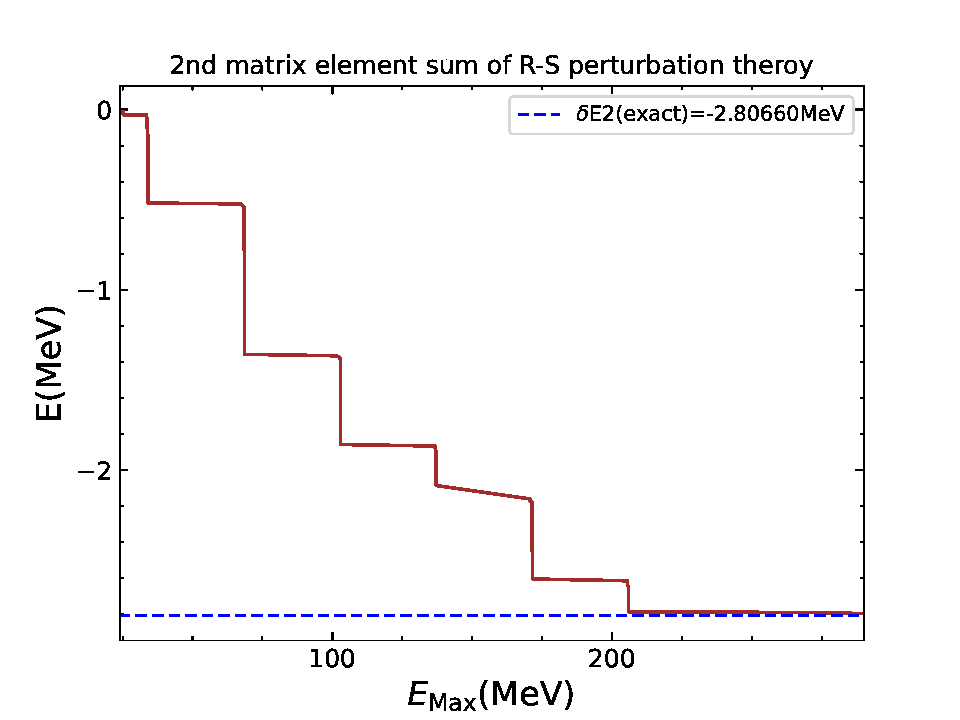}
\par\end{centering}
\caption{The second order perturbative correction of the deuteron binding energy in a $L=7$ box calculated with the Rayleigh-Schr{\"o}dinger perturbation theory.
The zeroth order Hamiltonian is $H_{\rm SU4}$ and the target Hamiltonian is $H_{\rm NLO}$ with the momentum cutoff $\Lambda=400$ MeV.
The summation over the excited states is limited with a maximal excitation energy $E_{\rm max}$.
The points represent the excited states.
The number in the box represents the exact value obtained from the contour integral.
\label{fig:RSperturbation}}
\end{figure}

For the ground state expansion we found a slightly abnormal convergence
pattern. The second order correction is much larger than the first
order correction. This point has been explained in Ref.\cite{PhysRevLett.128.242501}
by considering the symmetries of the non-perturbative and the target
Hamiltonian. As $H_{{\rm SU4}}$ has a much larger symmetry group
than $H_{{\rm NLO}}$, the eigen states of the former Hamiltonian
have additional quantum numbers which can be mixed by the latter Hamiltonian.
In terms of the conventional Rayleigh-Schr{\"o}dinger perturbation theory,
the correction term $V_{C}=H_{{\rm NLO}}-H_{{\rm SU4}}$ bridges different
irreducible representations of the SU4 group and induces non-zero
matrix elements between the ground state and the excited states. Such
matrix elements enter the second order correction via
\begin{equation}
\setlength\abovedisplayskip{18pt}
\setlength\belowdisplayskip{18pt}
E_{\lambda^2}=\sum_{n>0}\frac{|\langle\Psi_{0}|V_{C}|\Psi_{n}\rangle|^{2}}{E_{0}-E_{n}},\label{eq:RSperturbation}
\end{equation}
where $E_{n}$ and $|\Psi_{n}\rangle$ are the $n$-th eigenenergy and
eigenfunction, respectively. To quantitatively see the contribution
to the second order correction from each excited state, in Fig.~\ref{fig:RSperturbation}
we depict the energy correction Eq.~(\ref{eq:RSperturbation}) with
the summation over $n$ limited to energy levels with the excitation energy $E\leq E_{{\rm max}}$.
Each point denotes a single excited state.
We see that $E_{{\lambda^2}}$ converge slowly as $E_{{\rm max}}$
increases. Although the result finally converge to the exact value
for $E_{{\rm max}}>200$ MeV, the contributions from each excited
state are usually tiny and we have to include hundreds of them for
a precision calculation. Finally, for the second order correction
we found good agreement among the Rayleigh-Schr{\"o}dinger perturbation
theory, the Cauchy integral and the ptQMC method. Note that for mass
number $A\geq4$ the direct diagonalization is not possible and the
first two methods are not available. 

\begin{figure}
\begin{centering}
\vspace{-1em}
\includegraphics[width=1\columnwidth]{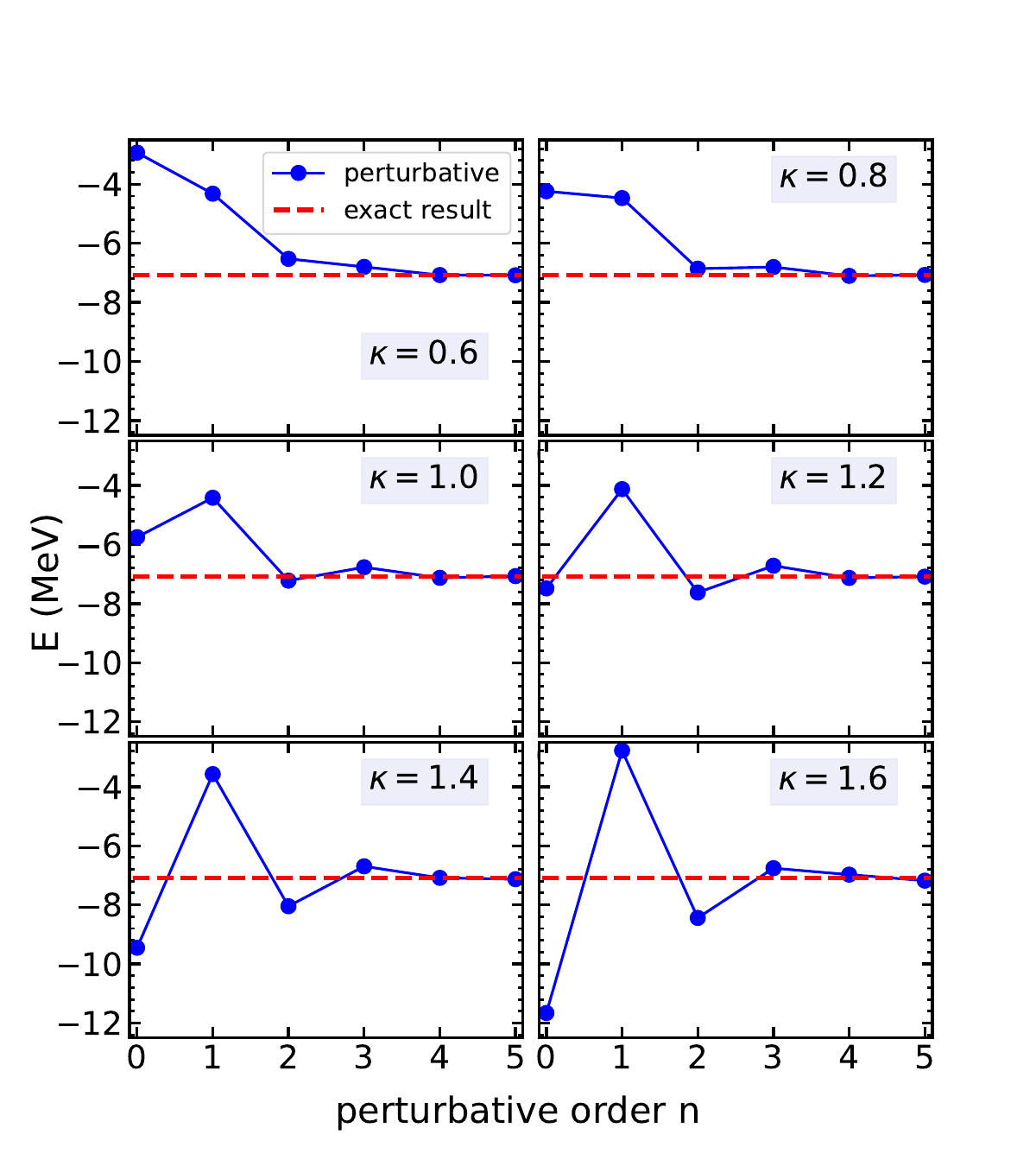}
\par\end{centering}\vspace{-0.9em}
\caption{The partial sums of the perturbative corrections up to each order for the ground state energy of the deuteron enclosed in a $L=7$ box. 
The number $\kappa$ denotes the multiplicative rescaling factor for the strength of the SU4 interaction $C_{\rm SU4}$ appearing in $H_{\rm SU4}$.
In each panel the dashed line represents the exact binding energy obtained from diagonalizing the target Hamiltonian $H_{\rm NLO}$ with the momentum cutoff $\Lambda=400$ MeV.
\label{fig:mudependence}}
\end{figure}

\begin{figure}
\begin{centering}
\vspace{0em}
\includegraphics[width=1\columnwidth]{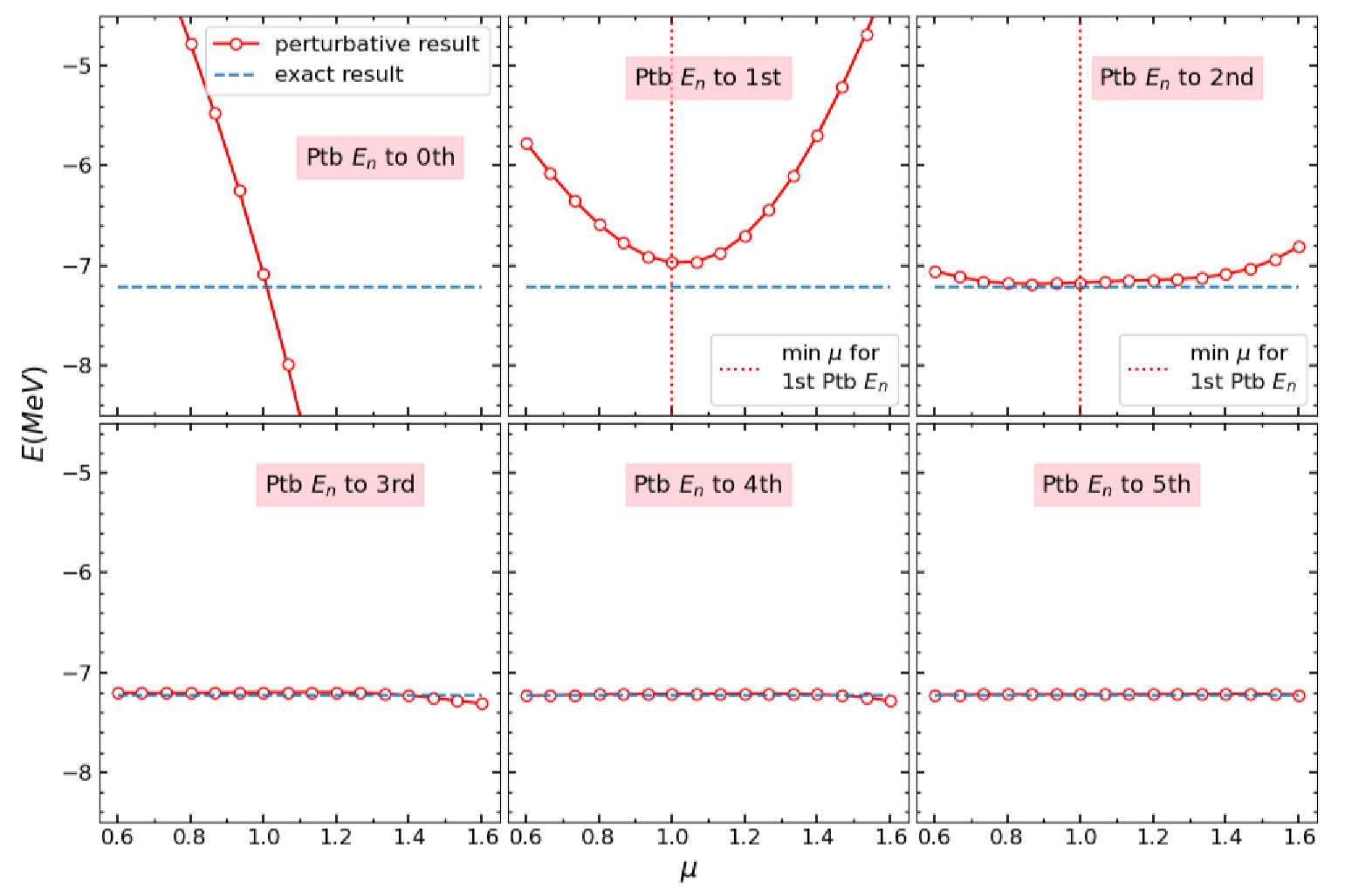}
\par\end{centering}
\vspace{-1em}
\caption{The same as Fig.~\ref{fig:mudependence}, but with the rescaling factor $\kappa$ as the horizontal axis.
Each panel corresponds to a partial sum up to a specific order.
The vertical dotted lines mark the location of the minimum for the first order energies.
\label{fig:mufunction}}
\end{figure}

In a practical ptQMC calculation for
many-body systems, the exact solution is usually unknown. Futhermore,
the perturbative corrections beyond the second order are difficult
to obtain. Thus it is challenging to estimate the uncertainty from
truncating the perturbative series. To this end we can repeat the
calculation with different zeroth order Hamiltonian and check
the corresponding variations of the results. For example, we can vary
the strength of the SU4 interaction and examine how the perturbative
series converge to the same result order by order.
Here we again take the ground state of the
deuteron to illustrate the idea. In Fig.~\ref{fig:mudependence} we
show the partial sums of the energy corrections calculated with different
rescaling factor $\kappa$ for the SU4 interaction. Here the non-perturbative
Hamiltonian $H_{{\rm SU4}}(\kappa)$ varies and the target Hamiltonian
is fixed to $H_{{\rm NLO}}$, with their difference $H_{{\rm N3LO}}-H_{{\rm SU4}}(\kappa)$
solved perturbatively. We see that for $0.6\leq\kappa\leq1.6$ the results
all converge to the exact eigenvalue of $H_{{\rm NLO}}$, whereas
the corrections of the first few orders vary significantly. In particular,
the zeroth order energy obtained as the eigenvalue of $H_{{\rm SU4}}(\kappa)$
is shifted from about $-$3 MeV to $-12$ MeV, such a wide uncertainty
interval shrinks quickly as we proceed to higher perturbative orders.
This observation is more clearly seen in Fig.~\ref{fig:mufunction},
where we depict the partial sums of the perturbative series as functions
of the rescaling parameter $\kappa$. We find that the energies beyond the second order are almost independent of
$\kappa$. 

The precision of the perturbative calculations can be further improved
by noticing the fact that the first order calculation is actually
variational. Up to this order the energy is just the expectation value
of the full Hamiltonian sanwiched between the zeroth order wave function.
We can find an optimized zeroth order Hamiltonian by requiring that
the first order energy is minimized against the parameters in $H_{0}$.
In the second panel of Fig.~\ref{fig:mufunction} we plot the optimal
value of $\kappa$ with a vertical dotted line. At this point the $H_{{\rm SU4}}(\kappa)$
ground state has the largest overlap with the $H_{{\rm NLO}}$ ground
state. In this case the essential part of $H_{{\rm NLO}}$ has been
absorbed in $H_{{\rm SU4}}(\kappa)$ and we expect that the perturbative
corrections are minimized. This conjecture can be verified by examining
the higher order corrections. In the third panel of Fig.~\ref{fig:mufunction}
we plot the same vertical line for the optimal $\kappa$. Clearly for
this values of $\kappa$ the second order energy is
almost equal to the exact energy with the higher order corrections negligible. 

\subsection{Solution of the N$^3$LO Hamiltonian}

In this section we perform the ptQMC calcualtion for the high-fidelity N$^3$LO chiral forces given by Eq.~(\ref{eq:N3LOexpansion}) and Table \ref{tab:Fitted-LECs-at}.
The method is described in Sec.~\ref{Sec:ptQMC}.
Here the resulting energies are given as a double expansion for the small parameters $\lambda$ and $\mu$ defined in Eq.~(\ref{eq:N3LOexpansion}).
As the term proportional to $\mu$ is at least one order smaller than that propotional to $\lambda$, we have assumed that the $\mathcal{O}(\mu^2)$ terms can be safely ommited. 
Now we examine this point numerically.

\begin{figure}
\begin{centering}
\includegraphics[width=1.0\columnwidth]{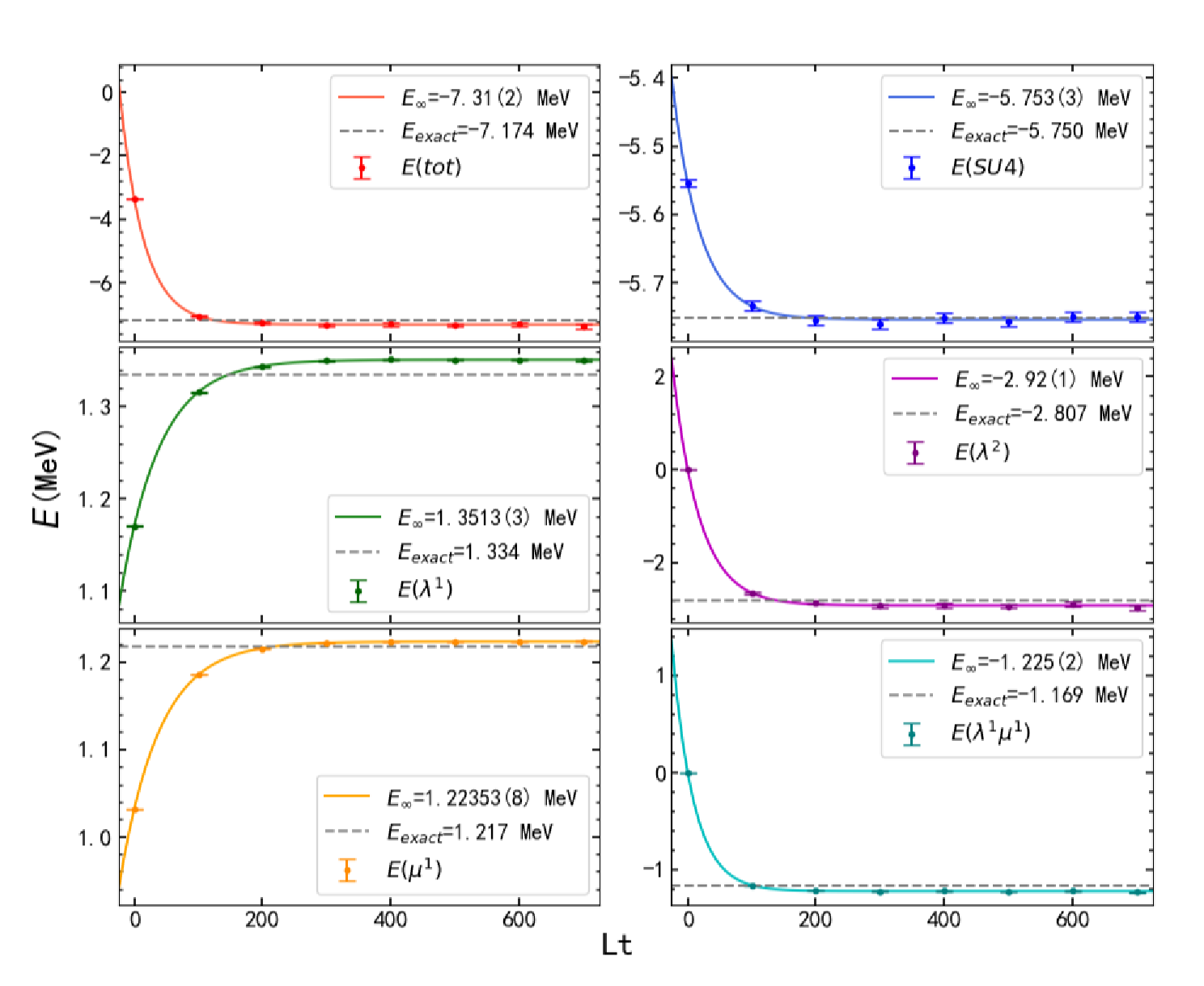}
\par\end{centering}
\caption{The perturbative corrections at each order for the ground state energy of the deuteron enclosed in a $L=7$ box.
$E_{\rm tot}$ is the sum of the other five components.
The zeroth order Hamiltonian is $H_{\rm SU4}$ and the target Hamiltonian is $H_{\rm N3LO}$ with the momentum cutoff $\Lambda=400$ MeV. We employ a double expansion corresponding to Eq.~(\ref{eq:N3LOexpansion}).
The symbols with errorbars represent the ptQMC results and the solid lines are fitted exponential curves for extrapolating to infinite projection time $L_t\rightarrow\infty$.
The horizontal dashed lines represent the exact values from the Cauchy contour integral.
On each panel the extrapolated and exact results from the contour integrals are given as $E_\infty$ and $E_{\rm exact}$, respectively.
\label{fig:RSperturbation-1-1-1-1}}
\end{figure}

In Fig.~\ref{fig:RSperturbation-1-1-1-1}  we display the evolution of  the various contributions of the deuteron binding energy against the number of time slices $L_t$.
For consistency we still use a small box $L=7$ in which the exact deuteron binding energy at N$^3$LO is $-$7.22 MeV.
We find that the convergence pattern is similar for different momentum cutoffs and only show the results with the momentum cutoff $\Lambda=400$ MeV.
Here the points with errorbars are all obtained with Monte Carlo simulations. For each curve we extraploate to $L_t\rightarrow\infty$. 
The converged values $E_\infty$ together with their uncertainties are given in the legends of each subplots.
These uncertainties are combined errors from both the statistical error and the $L_t$-extrapolation.
The total energy $E_{\rm tot}$ is the sum of all five terms $E_{\rm SU4}$, $E_{\lambda^1}$, $E_{\lambda^2}$, $E_{\mu^1}$ and $E_{\lambda^1\mu^1}$.
We see that all these energies converge rapidly for large $L_t$.
To further benchmark the partial perturbative contributions, We apply the Cauchy integral method introduced in Sec.~\ref{Sec:HNLOexpansion} to calculate the exact values of these terms.
The results are denoted as $E_{\rm exact}$ in each panel.
We find that the extrapolated results $E_{\infty}$ agree well with the exact values $E_{\rm exact}$.
The remaining discrepancies are mainly due to the finiteness of the temporal steps and can be further eliminated by the extrapolation $a_t\rightarrow 0$.
The Monte Carlo total energy $E_{\rm tot}$ converges to $-$7.31(2) MeV, while the sum of the exact values gives $-$7.174 MeV.
Note that these two values are both perturbative approximations and their difference reflects the precision of the imaginary time projection algorithm.
Comparing these results with the exact binding energy $-$7.22 MeV from sparse matrix diagonalization, we see immediately that the higher-order corrections like the $\mathcal{O}(\mu^2)$ terms are as small as 0.05 MeV, which justifies the ommision of such terms in ptQMC calculations.

One important observation is that the last two terms linear in $\mu$ approximately cancel each other.
This can be explained by noticing that the sum of the first three terms independent of $\mu$ is just the $\mathcal{O}(\lambda^2)$ approximation of the exact NLO energy $E_{\rm NLO}=-$7.08 MeV. 
For the deuteron we see that the exact N$^3$LO energy $E_{\rm N3LO}=-$7.22 MeV is close to that exact NLO energy, thus the $\mu$-dependent terms must be highly suppressed.
Such a suppression is due to the EFT power counting and must be manifested in both the exact and perturbative calculations.
This situation is similar to the cases of perturbative calculations with special symmetries. 
Although the perturbative expansion may break certain symmetries and induce large symmetry-breaking terms, in the final result the various contributions should cancel each other to recover the symmetry.
Here the power counting tells us that the difference between $E_{\rm N3LO}$ and $E_{\rm NLO}$ is definitely small.
However, for a perturbative expansion around $H_{\rm SU4}$, the operator $H_{\rm N3LO}-H_{\rm NLO}$ is inserted into the wave function of $H_{\rm SU4}$ rather than that of $H_{\rm NLO}$, resulting in a large contribution of about 1~MeV.
Such a spurious correction can be remedied by improving the wave functions using the NLO Hamiltonian, which gives a term proportional to both $\lambda$ and $\mu$. 
Finally the summation of the $\mathcal{O}(\mu)$ and $\mathcal{O}(\lambda\mu)$ terms amounts to the small difference $E_{\rm N3LO} - E_{\rm NLO}$.
Thus the cancellation is not a coincidence and can be considered as an indication of the effectiveness of the perturbation theory.
As a consequence, in general ptQMC calculations, we must check whether the sum of the contributions from the individual term agree with the overall estimation from the power counting scheme.

Next we turn to the many-body calculations with $A\geq3$.
In this case the terms $V_{\rm CSB}$ and $V_{\rm GIR}$ become essential.
We note that the purpose of this work is to demonstrate the capability of the algorithm.
To this end, we can either compare the perturbative results with the exact solutions as in the case of the deuteron, or contrast the results with the experiments whenever no exact solution is available. 
Unfortunately, for $A\geq4$ there is no exact solution for the lattice Hamiltonian.
Further, as we have not determined the three-body forces in constructing the chiral forces, we do not expect a accurate reproduction of the experimental values.
Nevertheless, there are other phenomena we can exploit to benchmark our algorithm.
For example, it was long known that the binding energies of $^3$H and $^4$He calculated with different two-body interactions form a straight line passing through the experimental point.
It is thus important to check whether our algorithm can reproduce such a Tjon band~\cite{PhysRevC.20.340,PhysLetterB.607.254,EPJA54-121}. 
This examination is a necessary but not a sufficient condition for the correctness of the ptQMC method.
To obtain more data points, here we employ chiral interactions with multiple momentum cutoffs.

\begin{figure}
\begin{centering}
\includegraphics[width=1\columnwidth]{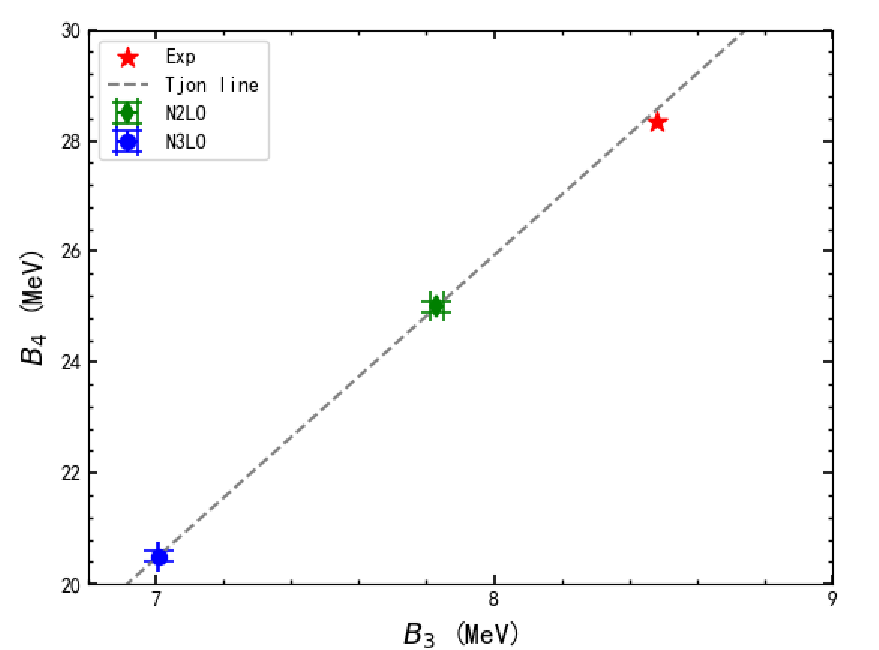}
\par\end{centering}
\caption{The binding energies $B(^3$H) versus $B(^4$He) calculated with various nuclear forces and many-body methods.
The triangles, squares and circles denote the results calculated with the Hamiltonians Eq.~(\ref{eq:DefineHamiltonian}) parametrized with the LECs in Table~\ref{tab:Fitted-LECs-at}.
The open and full symbols represent the NLO and N$^3$LO results, respectively.
The momentum cutoff $\Lambda$ is given in the parentheses.
The $^3$H and $^4$He energies are calculated using the Lanczos method and ptQMC method, respectively.
The results have been extrapolated to $L\rightarrow\infty$ to eliminate the finite volume effects.
The results with other interactions are taken from Ref.~\cite{PhysRevLett.85.944,PhysRevC.103.054001}.
The experimental values are marked as a red star.
An empirical Tjon band is depicted to guide the eyes.
\label{fig:RSperturbation-1-1-1-1-1}}
\end{figure}

In Fig.~\ref{fig:RSperturbation-1-1-1-1-1} we plot the binding energies of $^3$H and $^4$He calculated with the chiral interactions defined in Eq.~(\ref{eq:N3LOexpansion}) and parametrized with Table \ref{tab:Fitted-LECs-at}.
The results with $\Lambda=300$, $350$ and $400$ MeV are denoted as triangle, square and circle, respectively.
The results at NLO and N$^3$LO are displayed as open and full symbols, respectively.
The experimental values are shown as a red star.
Note that we treat $^3$H and $^4$He differently.
We always calculate the exact $^3$H binding energies using the Lanczos method, while we solve the $^4$He binding energies with the ptQMC method up to the second order as described in the previous sections.
The advantage of this prescription is that the exact $^4$He energies can be inferred from its correlation with the $^3$H energies.
If the resulting points is far off from the Tjon line, we know that the perturbative calculations have missed certain essential elements.
To improve the precision, for each point we have made calculations with different box size $L=11, 13, 15, 17$ and extrapolate to the infinite volume limit $L\rightarrow\infty$.
The uncertainties from both the Monte Carlo statistical errors and extrapolation errors are smaller than the symbols and not shown here.
In Fig.~\ref{fig:RSperturbation-1-1-1-1-1} we find that all the results together with the experimental point can be covered by a narrow pink Tjon band.
For reference we also plot the points calculated with the nuclear forces constructed in the continuum space and solved with various many-body methods. 
We see that the Tjon line generated by our calculations is slightly lower than the version from the continuum nuclear forces.
We attribute this difference to the usage of the lattice regulator.

In Table ~\ref{Tab:4He_partial_E_at_diff_lambda} we show the total $^4$He energies and the partial perturbative corrections. 
We show both the NLO and N$^3$LO results with different momentum cutoffs.
For simplicity we fix the box size to $L=15$ and omit the infinite volume extrapolations. 
For all calculations, we start from the same SU4 interaction Eq.~(\ref{eq:HSU4}) and obtain the same values for $E_{\rm SU4}$.
The tiny differences are due to the Monte Carlo statistical errors.
For fixed value of $\Lambda$, the corrections $E_{\lambda^1}$ and $E_{\lambda^2}$ are only determined by $H_{\rm NLO}-H_{\rm SU4}$ and assume the same values for NLO and N$^3$LO calculations.
Compared with the N$^2$LO calculations including the three-body forces~\cite{PhysRevLett.128.242501}, here we only focus on the two-body forces and employ a modified SU4 interaction designed for smaller lattice spacing, thus the results at each order are different. 
For example, the $E_{\lambda^1}$ corrections are much larger than that found in Ref.~\cite{PhysRevLett.128.242501}. Nevertheless, the partial sums up to the $\lambda^2$ order are all about $-$25~MeV, which fits well to what we expect for the two-body force contributions.
The terms $E_{\mu^1}$ and $E_{\lambda^1\mu^1}$ only appear in N$^3$LO calculations.
In comparison with the N$^3$LO calculation for the deuteron, a remarkable difference is that $E_{\rm N3LO} - E_{\rm NLO}$ becomes much larger in this case and the $\mu$-dependent terms do not cancel up.
The reason is that we have not included the three-body forces which are naturally generated when we slide the cutoffs or introduce more higher-order operators. 
For the deuteron case, the difference $H_{\rm N3LO} - H_{\rm NLO}$ can be a negligibly small term as both Hamiltonians generate almost the same two-body phase shifts. 
However, for systems with $A\geq3$, their difference could become more significant and this discrepancy can be remedied by including three-body forces calibrated within three-body systems.
This explains why the $\mu$-dependent terms are sizable for $^3$H and $^4$He nuclei.
Nevertheless, without the three-body forces the $\mu$-dependent corrections are still small compared with the total binding energy and we expect that the linear approximation keeps valid here.
The convergence pattern we found for the deuteron can be recovered in future many-body calculations when the three-body forces are faithfully included.



\begin{table*}[]
    \centering
    \caption{The perturbative corrections at each order for the ground state energy of $^4$He enclosed in a $L=15$ box. 
    $E_{\rm tot}$ is the sum of the other five components. 
    The zeorth order Hamiltonian is $H_{\rm SU4}$ and the target Hamiltonian is $H_{\rm N3LO}$. We employ a double expansion corresponding to Eq.~(\ref{eq:N3LOexpansion}).
    The numbers in the parentheses denote the momentum cutoff $\Lambda$.
    The uncertainties are the statistical errors from the Monte Carlo simulation.
    All units are in MeV.}
    {\small 
  \setlength{\tabcolsep}{22pt} 
  \begin{tabular}{c|cccccc}
    \hline
    & E$_{\text{tot}}$ & E$_{\text{SU4}}$ & E$_{\lambda^1}$ & E$_{\lambda^2}$ & E$_{\mu^1}$ & E$_{\lambda^1\mu^1}$ \\
    \hline
    NLO (300)     & $-25.5(1)$ & $-23.36(4)$ & $3.0$    & $-5.1(1)$  & 0    & 0    \\
    N$^3$LO (300) & $-23.4(1)$ & $-23.36(4)$ & $3.0$    & $-5.1(1)$  & $2.7$ & $-0.6$ \\
    NLO (350)     & $-26.1(1)$ & $-23.36(5)$ & $3.2$    & $-5.9(1)$  & 0    & 0    \\
    N$^3$LO (350) & $-22.9(1)$ & $-23.36(5)$ & $3.2$    & $-5.9(1)$  & $4.4$ & $-1.2$ \\
    NLO (400)     & $-25.3(1)$ & $-23.37(4)$ & $7.0$    & $-9.0(1)$  & 0    & 0    \\
    N$^3$LO (400) & $-20.9(1)$ & $-23.37(4)$ & $7.0$    & $-9.0(1)$  & $6.3$ & $-1.9$ \\
    \hline
  \end{tabular}%
}
\label{Tab:4He_partial_E_at_diff_lambda}
\end{table*}

\section{Summary and perspective}

The sign problem is one of the most notorious obstacles that prevent
us from simulating many quantum many-body systems. For systems involving multiple
fermions such as the atomic nuclei, a partial solution of the sign
problem means a much deeper understanding for a large variety of phenomena.
Fortunately, the nature is usually close to some prototypical models
that are free from the sign problem and solvable with the Monte Carlo
methods. This is especially true for nuclear physics. The nuclear
force is very close to a Wigner SU4 limit, in which the interaction
is independent of the spin and isospin. In this limit the even-even
nuclei can be simulated without the sign problem. Although the high-precision
nuclear forces usually assume very complicated structures consisting of tens
of different contact terms and pion exchange potentials, all the terms
invoving spin or isospin will be suppressed due to this approximate
symmetry. This fact inspires us to develop a perturbation theory in
the framework of the projection MC calculation. This perturbative quantum Monte Carlo (ptQMC) method turns to be very efficient for a realistic next-to-leading-order (NLO) chiral force~\cite{ref11_in_PRL128-242501}.
We collect all the higher order terms and one-pion-exchange-potential
into the perturbative Hamiltonian and calculate their contribution
to the energy up to second order. This ptQMC method allows us to make
MC calculations with large projection time with a polynomial computational cost.

In this work we perform the first ptQMC calculations with the next-to-next-to-next-to leading order (N$^3$LO) chiral forces.
The challenges we faced are two-folds.
Firstly, it is difficult to construct the high-fidelity nuclear forces on the lattice without introducing large lattice artifacts.
Here we use the same recipe as in Ref.~\cite{ref11_in_PRL128-242501} that smearing the interactions with a relatively low momentum cutoff.
Such a cutoff smoothly screens the lattice artifacts and preserves the symmetries.
The remaining broken symmetries such as the Galilean invarance can be restored by additional counter terms in the Hamiltonian.
Secondly, at N$^3$LO the number of independent operators is so large that a direct application of the ptQMC method is not realistic.
Here we propose a calculation scheme that treats the corrections differently based on the hierarchical structure of the chiral forces.
We take the deuteron as an example to show the general pattern of the perturbative series in this method.
Note that in our work the perturbative calculation is only viewed as a numerical tool for solving the given chiral Hamiltonian.
Our target is always the exact value of the eigen energies and has nothing to do with the construction of the chiral forces.
Thus in principle our method can also be applied to general nuclear interactions other than the chiral forces.

Recently it was proposed that the nuclear physics might be expressed
as a perturbative expansion near a unitary limit, where nuclear force
satisfies the Wigner SU4 symmetry and the $S$-wave scattering length
goes to infinity\cite{PhysRevLett.118.202501}. A relatively weaker statement only
requries the SU4 symmetry. These conjectures have
been numerically examined for few-body systems with $A\leq4$ based
on the pionless EFT. However, so far there is no order-by-order check
for the convergence of such perturbative series in many-body calculations.
The
results presented here will not only provide a powerful tool for solving
the sign problem, but also serves as a framework for understanding
the various universalities observed in nuclear \textit{ab initio}
calculations. For example, the Tjon line, Phillips line and Coester
band are all consequences of the proximity of the nuclear force to
the unitary limit. We expect that the higher order perturbative calculations
can bridge the gap between the Wigner SU4 limit or unitary limit
which are simpler to understand and the real world in which the nuclear
force takes a seemingly more complicated form.

Finally, it is also possible to generalize the method in this paper
to compute third or even higher order perturbative corrections within
a polynomial time. For converging series these contributions will
decrease quickly and be submerged by the statistical noise at certain
orders, which provides a natural cutoff. Future works will include
analyzing the higher order terms, examining the convergence radii
against some parameters such as the momentum cutoff and applying the
method for other systems plagued by the sign problem.

\section{Acknowledgment}
We are grateful for discussions with members of the Nuclear Lattice Effective Field Theory Collaboration.
This work has been supported by 
 NSAF No.U2330401 and National
 Natural Science Foundation of China with Grant No.12275259.

\setcitestyle{sort&compress}
\bibliographystyle{unsrt}

\end{document}